\documentclass[trans]{IEEEtran}
\usepackage{amsmath,amsthm}
\usepackage{graphics} 
\usepackage{epsfig,epstopdf} 
\usepackage[export]{adjustbox}
\usepackage[dvipsnames]{xcolor}
\usepackage{amsmath,bigints,bbm} 
\usepackage{amssymb}  
\usepackage[T1]{fontenc}
\usepackage[utf8]{inputenc}
\usepackage{authblk}
\usepackage{cite}
\usepackage{cleveref}
\usepackage{caption}
\usepackage{subcaption}
\usepackage{mathtools}
\usepackage{accents}

\DeclareFontFamily{U}{mathx}{\hyphenchar\font45}
\DeclareFontShape{U}{mathx}{m}{n}{
	<5> <6> <7> <8> <9> <10>
	<10.95> <12> <14.4> <17.28> <20.74> <24.88>
	mathx10
}{}
\DeclareSymbolFont{mathx}{U}{mathx}{m}{n}
\DeclareFontSubstitution{U}{mathx}{m}{n}
\DeclareMathAccent{\widecheck}{0}{mathx}{"71}
\DeclareMathAccent{\wideparen}{0}{mathx}{"75}

\newtheorem{proposition}{Proposition}

\newtheorem{assumption}{Assumption}

\newtheorem{definition}{Definition}
\crefrangelabelformat{equation}{(#3#1#4--#5#2#6)}
\crefname{equation}{Eq.}{Eqs.}
\Crefname{equation}{Equation}{Equations}

\newtheoremstyle{problemstyle}  
        {3pt}                                               
        {3pt}                                               
        {\normalfont}                               
        {}                                                  
        {\bfseries\itshape}                 
        {\normalfont\bfseries:}         
        {.5em}                                          
        {}                                                  
\theoremstyle{problemstyle}

\usepackage{algorithm}
\usepackage{algorithmic}
\usepackage{tikz}
\usetikzlibrary{positioning}

\usepackage{setspace}	

\begin{document}
\title{ The Role of UAV-IoT Networks in Future Wildfire Detection }

\author{  
\IEEEauthorblockN{Osama M. Bushnaq, {\em Student Member, IEEE}, Anas Chaaban, {\em Senior Member, IEEE}, 
	 and Tareq Y. Al-Naffouri, {\em  Member, IEEE}.} 
\thanks{O. M. Bushnaq, and T. Y. Al-Naffouri  are with Computer, Electrical and Mathematical Sciences and Engineering (CEMSE) Division of King Abdullah University of Science and Technology (KAUST), Thuwal, KSA. (E-mails: {osama.bushnaq@kaust.edu.sa}, and {tareq.alnaffouri@kaust.edu.sa})  

A. Chaaban is with the School of Engineering, the University of British Columbia, Kelowna V1V 1V7, BC, Canada (E-mail: {achaab01@mail.ubc.ca})
}
}

\maketitle
\begin{abstract}
	
	The challenge of wildfire management and detection is recently gaining increased attention due to the increased severity and frequency of wildfires worldwide. Popular fire detection techniques such as satellite imaging and remote camera-based sensing suffer from late detection and low reliability while early wildfire detection is a key to prevent massive fires. In this paper, we propose a novel wildfire detection solution based on unmanned aerial vehicles assisted Internet of things (UAV-IoT) networks. The main objective is to (1) study the performance and reliability of the UAV-IoT networks for wildfire detection and (2) present a guideline to optimize the UAV-IoT network to improve fire detection probability under limited budgets. We focus on optimizing the IoT devices' density and number of UAVs covering the forest area such that a lower bound of the wildfires detection probability is maximized within a limited time and budget. At any time after the fire ignition, the IoT devices within a limited distance from the fire can detect it. These IoT devices can then report their measurements only when the UAV is nearby. Discrete-time Markov chain (DTMC) analysis is utilized to compute the fire detection probability at discrete time. Before declaring fire detection, a validation state is designed to account for IoT devices' practical limitations such as miss-detection and false alarm probabilities. Numerical results suggest that given enough system budget, the UAV-IoT based fire detection can offer a faster and more reliable wildfire detection solution than the state of the art satellite imaging techniques.
	
\end{abstract}
\section{Introduction}

In the last few years, the number, frequency, and severity  of wildfires have increased dramatically worldwide, significantly impacting countries economies, ecosystem\footnote{In some cases, forest fires can be beneficial to maintain a healthy and diverse ecosystem.}, and communities. For instance, an average of 2.5 million hectares (ha) are burnt only in Canadian forests every year, which costs around 370 million to 740 million US dollars per year \cite{2018CanadaForestAnnualReport,hope2016wildfire}. The risk of wildfires is expected to increase in the near future, see \cite{hope2016wildfire, 2019TheBurningIssue} for more insights and statistics. 

The management of wildfires forms a significant challenge where early fire detection is key. Current wildfire detection methods such as satellite imaging and infrared cameras are not reliable especially under cloudy weather conditions. In order to detect wildfires before getting out of control, we can make use of IoT networks, which can connect a massive number of simple-structured, self-powered and cheap IoT sensors \cite{IoT2015,IoT2013}. While IoT networks are expected to support 1 million IoT devices per km$^2$ by 2025 \cite{Qualcomm}, the lack of infrastructure over forests and the limited IoT devices’ power and complexity make data aggregation unattainable using standard IoT networks. 

To solve this problem, UAVs can be used. UAVs can support increased data rates and reliability demands for cellular communication networks \cite{Zeng2016May}. In addition to this, UAVs offer the advantage of flexibility and decreased costs, which makes them suitable to reach dangerous and remote areas for disasters recovery. Therefore, many recent studies have suggested UAV-IoT networks to manage natural disasters \cite{Bushnaq2019,LTE_Sky,Public_UAV}. 

In this paper, we propose a new wildfire detection method based on Unmanned Aerial Vehicle aided Internet of Things (UAV-IoT) network. The aim of the study is to evaluate the reliability of the UAV-IoT networks in detecting wildfires within a limited period of time. Further, we study the optimal density of IoT devices and the number of UAVs such that a lower bound of the probability of fire detection is maximized under a limited system budget. To the best of the authors knowledge, there is no work which considers and studies UAV-IoT networks specifically for wildfire detection. Before elaborating this solution, we discuss some background related to this application area.

\subsection{Background}
Studying wildfire detection relies on two important ingredients: (i) fire spread models and (ii) fire detection. We briefly discuss these two ingredients next. 

Fire spread is a dynamic process which depends on environmental variables such as wind speed, moisture content, fuel type and density, ground slope, etc. Developing an accurate fire spread model which can predict fire size and shape over time is an ongoing research challenge. Wildfire spread modeling can be categorized into physics-based \cite{physics-based2007, physics-based2010} and experiment-based approaches \cite{Boychuk2009,sim_based2013}. Physics-based models suffer from oversimplification of the complex forest environments, while empirical models suffer from the lack of accurate experimental data over the burnt forests which can be utilized to validate the mathematical model. A popular empirical approach employing Markov stochastic process to model the fire spread is discussed in \cite{Boychuk2009}. We use this latter empirical approach in this paper. 

The main methods to detect wildfires today can be categorized into satellite imaging, remote sensing, and wireless sensor network (WSN) detection \cite{Alkhatib2014}. These methods are discussed below.

\subsubsection{Satellite Imaging}

Satellite based forest monitoring and fire detection is the most popular approach today due to its low cost. Advanced geostationary satellite systems such as the Advanced Very-High-Resolution Radiometer (AVHRR) can take images with as low spatial and temporal resolutions as one km$^2$ and about ten minutes, respectively \cite{Satellite2016, Satellite2018, Satellite2019}. Despite this resolution, fire detection at early stages is not possible using this method, since the fire area has to be already multiple km$^2$ to be observable, and some fires can spread vastly within few minutes. Further, the quality of satellite images is highly dependent on weather conditions.

\subsubsection{Remote Sensing} Mounting surveillance cameras or infrared-sensors on a ground tower, UAV or high altitude platform (HAP) is another popular wildfire detection method. With the advancement of camera technology, Artificial Intelligence (AI), computer vision and image recognition, this approach has gained more interest recently \cite{camera2018, camera2018_CNN, camera2017}. While installing ground surveillance stations is expensive, mounting cameras on UAVs is more promising. A survey for fire monitoring, detection, and fighting techniques using UAV is presented in \cite{Yuan2015}. The reliability of this approach decreases under cloudy weather conditions. Another challenge to this method is the limited UAV battery which is drained by the camera sensor and the complex AI processing. While the UAV images may not be able to cover a wide forest area, HAPs constitute a compromise between the gains and drawbacks of satellite fire detection and UAV remote sensing \cite{Allison2016HAP, Isabelle2020HAP}. After detection at the UAV, it transmits its result to a central station. UAV-assisted communication is essential to report fire detection, rescue communities close to the fire and keep track of the fire evolution over time \cite{selim2018}. Spectrum sharing for UAVs at the emergency relief and fire diagnosis phase are discussed in \cite{shamsoshoara2019autonomous}.

\subsubsection{WSN detection}
Wireless sensor networks offer another technique for wildfire detection. Although not as popular as satellite imaging and remote sensing, WSN fire detection has attracted more research recently due to the enhanced efficiency and reduced costs of the WSNs. Sensing data such as temperature, smoke, moisture content, etc. is not only useful for wildfire detection, but also offers big amounts of data for forest monitoring. Sensors' measurements are essential to predict the fire behavior and diagnose its impact \cite{Yuan2015}. The main challenge of such systems is the limited sensor power which is needed to transmit its measurements to relatively far sensors/access-points. In \cite{UAV-IoT2008} the coexistence of WSN and remote sensing from UAV is suggested to enhance detection reliability. However, the interaction between the UAV and the WSN was not proposed in \cite{UAV-IoT2008}. 

In the following subsection, we summarize the contribution of this paper.

\subsection{Contributions}

In this paper, we propose a novel wildfire detection technique based on UAV-IoT networks. The main objective is to (1) study the performance and reliability of the UAV-IoT networks for wildfire detection and (2) present a guideline to optimize the UAV-IoT network to improve fire detection probability under a limited budget. Although there are several parameters affecting the system reliability, we focus on optimizing the IoT devices density and number of UAV covering the forest area. 

For the sake of simplicity, we assume a simplistic fire spread model, wherein fire evolves in a circular shape at a fixed speed, determined based on statistics of the environment of interest.\footnote{While this serves the purpose of a worst-case analysis, a more sophisticated model may be considered in simulation.} At any time after the fire ignition, the IoT devices within a limited distance from the fire are able to detect it. These IoT devices can then report their measurements when a patrolling UAV is within transmission range. Markov analysis is utilized to compute the fire detection probability at discrete time steps starting from the fire ignition time. Before declaring fire detection, a validation state is designed to account for IoT devices' miss-detection and false alarm probabilities. We then present insightful figures for the wildfire detection against several system parameters. 

The rest of the paper is organized as follows. In Section~\ref{sec:system_model}, we present the system model and the main assumptions. In Section~\ref{sec:problem_stat}, we discuss two problem statements; wildfire detection probability maximization and wildfire losses minimization. The detection performance analysis is presented in Section~\ref{sec:performance_analysis}, and design and performance insights are discussed in Section~\ref{sec:insigts}. Finally, numerical results are provided in Section~\ref{sec:numerical_results} before concluding the paper in Section~\ref{sec:conclusion}.

\section{System Model}
\label{sec:system_model}

Consider a large forest of area $A$ which we would like to protect against wildfires. A massive number $N_s$ of low cost, simple-structured and self-powered IoT devices (sensors with limited storage, processing and communication capabilities) are distributed at random locations over the area $A$ to detect fires, with a density of $\lambda_s = N_s/ A$ devices/km$^2$. We assume that each sensor is capable of detecting a fire at a distance of $d_s$ meters, and sets a flag to 1 or 0 if a fire or no fire is detected, respectively. 

Using such low-cost self-powered devices, complex routing algorithms and long-range transmission from the sensors to a fixed access point are prohibitive. Instead, sensors transmit their flags to a number $N_u$ of UAVs that hover over the forest and collect data from nearby sensors before traveling to a new location to collect data from another group of sensors. While UAVs would hover over deterministic paths in practice, we assume that they travel randomly in this work for tractability and generality, bearing in mind that results can only improve if optimized paths are used as in \cite{Julian2019DeepReinforcementLearning}. Each UAV collects an average of $N$ sensors' signals at each hovering location and enters a verification mode if it receives at least $M\leq N$ positive flags (i.e. a binary signal indicating a fire detection at the IoT device) at that location. In this case, the UAV spends an average of $T_{\rm vrf}$ sec. to verify the fire alarm raised by the collected observations. This verification can be done by sending photos to a central unit or by collecting more observations from neighboring regions. We assume that the UAV is capable of making a robust decision with negligible miss-detection and false alarm after this verification phase. Then, the system goes back to the normal search phase if the result is negative (no fire), or the system sends emergency signals to the fire fighting station otherwise. We elaborate on the system model below.

\subsection{Fire Spread Model}

Fire spread is a dynamic process which depends on environmental variables such as wind speed, moisture content, fuel type and density, ground slope, etc. Accurate fire spread modeling is fundamental to evaluate fire size and shape, and therefore help fire fighting teams assess and predict the danger level and the cost of slow/fast response. A popular approach is to utilize a Markov stochastic process to model the fire spread as in \cite{Boychuk2009}. 

In this model, the forest is divided into a $2D$ grid. Each block on fire spreads the fire to the neighboring blocks with some probability depending on environmental parameters such as fuel type, wind speed, moisture content, etc. \cite{Boychuk2009,Finney2011AMF,CRUZ201316}. As a result, the fire rate of spread (ROS) is evaluated in different directions and the shape of the fire is estimated at different times. To simplify the analysis in this paper, we make the following assumption.


{\assumption  We assume that all blocks have the same environmental properties and ignore the wind effect. As a result, a circular fire shape is formed with probability one in the long run. \label{ass:circular_fire}}
Although this assumption will not likely be satisfied in practice, it can be used as a means to assess worst-case performance. For a fixed fire size, a circular fire has the smallest perimeter. Hence, a circular fire has the smallest number of detecting sensors close to its front-line (i.e., its perimeter). Hence, if a system achieves good performance in detecting a circular fire, it will achieve equal or better performance in detecting a non circular fire. Next, we define forest area under fire at discrete time steps of duration $T$ sec. as follows.
{\definition Given that, without loss of generality, the fire starts at the origin, and taking Assumption~\ref{ass:circular_fire} into account, the fire at time step $k$ is spread over a disk defined in polar coordinates as,
\begin{align}
\mathcal{B}_f[k] = \{ (r,\theta): r \in [0, R_f[k]], \, \theta \in [0, 2\pi] \},
\end{align}
where $R_f[k] = vTk $ is the radius of the fire at time step $k$ and where $v$ represents the fire ROS.  
}

\subsection{Sensor detection model}
An IoT sensor can detect the fire based on environmental variations such as temperature, smoke, etc. measured at the IoT device location. The environmental variations $\delta(d)$ measured at an IoT device at distance $d$ from the fire front-line can be written as $\delta_n + \delta_f(d)$ where $\delta_n$ is the nature-induced component and $\delta_f(d)$ is the fire-induced one.

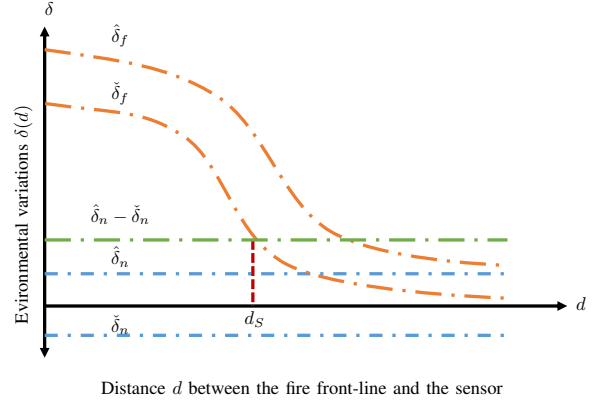
\begin{figure}
	\centering
	\begin{tikzpicture}[thick,scale=0.7, every node/.style={scale=0.7}]
	\draw (0, .7) node[inner sep=0] 
	{\includegraphics[trim={9cm 5cm  10cm 4cm},clip, width=1.3 \linewidth]{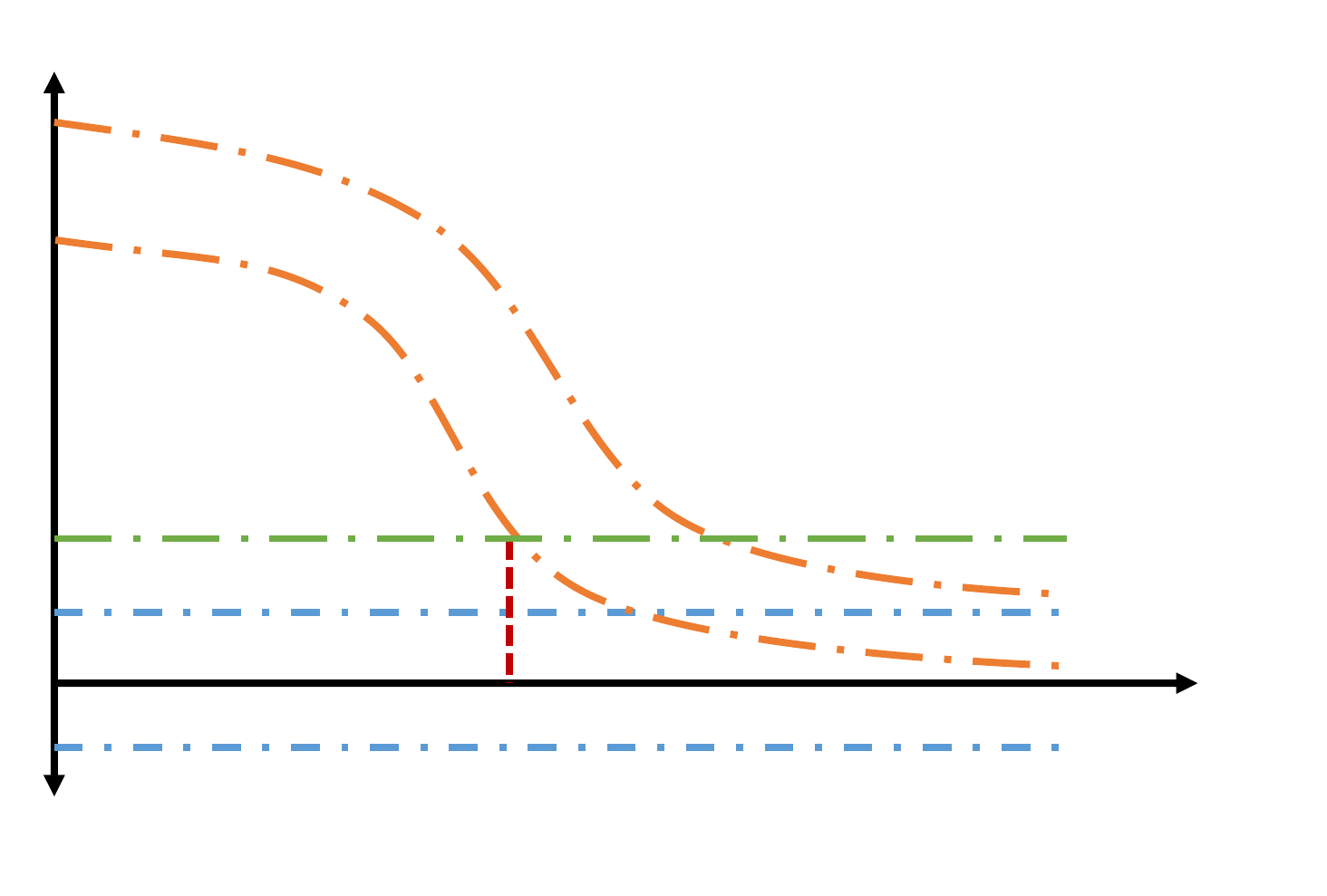} };
	\draw (-5.2,4.3) node {{$\delta$}};
	\draw (4.9,-1.3) node {{$d$}};
	\draw (-1.3,-1.6) node {{$d_S$}};
	
	\draw (-3.85,.4) node {{$\hat{\delta}_n -\check{\delta}_n$}};
	\draw (-3.85,-.4) node {{$\hat{\delta}_n$}};
	\draw (-3.85,-1.7) node {{$\check{\delta}_n$}};
	\draw (-3.85,3.8) node {{$\hat{\delta}_f$}};
	\draw (-3.85,2.7) node {{$\check{\delta}_f$}};

	\draw (-.4,-2.9) node {{Distance $d$ between the fire front-line and the sensor}};	
	\node [xshift=-6cm, yshift=.4cm, rotate=90,anchor=north] {Evironmental variations $\delta(d)$};
	
    
%
	\end{tikzpicture}
	\caption{A qualitative plot for the upper and lower bounds of the nature and fire-induced environmental variation at distance $d$ from the fire front-line.} \label{fig:env_par_var}
\end{figure}

We model $\delta_n$ and $\delta_f(d)$ as bounded random processes as shown in Fig.~\ref{fig:env_par_var} where $\hat{\delta_n}$ $\left(\check{\delta_n}\right)$ is the upper (lower) bound for the nature-induced variations and $\hat{\delta_f}(d)$ $\left( \check{\delta_f} (d)\right)$ is the upper (lower) bound for the fire-induced variations at distance $d$ from the fire front-line. Note that Fig.~\ref{fig:env_par_var} is a qualitative plot, that depicts the fact that $\delta_n$ is independent of $d$ and that $\delta_f$ decreases as $d$ increases. The manner of this decrease depends on the forest fuel type, ground slope, wind speed, etc. At the IoT device location, the IoT sensors indicate that there is a fire nearby by raising a flag if $\delta (d) > \hat{\delta}_n$, where $\hat{\delta}_n$ is known a priori at the forest geographical region\footnote{The nature-induced variation bounds can be sensed and adjusted by the IoT devices in a periodic fashion, or transmitted to the IoT devices from UAVs.}.

As shown in Fig.~\ref{fig:env_par_var}, a sensor at distance less than $d_s$ from the fire front-line with $\check{\delta_f}(d) > \hat{\delta}_n - \check{\delta}_n $ can detect the fire with probability one since,
\begin{align*}
\delta (d) &= \delta_f(d) +\delta_n \geq \check{\delta}_f (d) +\check{\delta}_n,\\
&> \hat{\delta}_n - \check{\delta}_n  + \check{\delta}_n = \hat{\delta}_n, \quad \forall d \leq d_s.
\end{align*} 
Sensors at distance $d\geq d_s$ from the fire may detect the fire with some probability. We focus on studying the worst case scenario by assuming sensors do not detect the fire at $d\geq d_s$. Based on this, we define the sensor detection ring as follows.

{\definition The IoT sensor detection ring is the set of points outside the fire front-line, within which sensors can detect the fire with probability one at time step $k$ (see Fig.~\ref{fig:system_model}). This is defined as,
\begin{align}
\mathcal{B}_s[k] = \{ (r,\theta): r \in [R_f[k],R_s[k] ], \quad \theta \in [0,2\pi]  \},
\end{align}	
where $R_{s}[k] = R_{f}[k]+d_s$.}
We assume that IoT devices inside the fire circle are damaged. Due to sensors' quality limitations in practice, sensors measurements and decisions are subject to sensing error $\epsilon_s$ that is independent of $d$. This leads to an erroneous detection outcome at the sensor level with probability $\epsilon_s$.

\subsection{IoT devices and UAVs Setup}
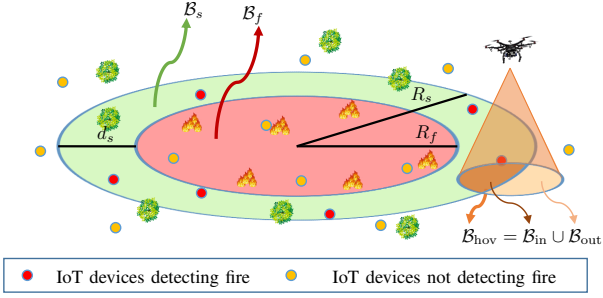
\begin{figure}
	\centering
	\begin{tikzpicture}[thick,scale=0.7, every node/.style={scale=0.7}]
	\draw (0, .7) node[inner sep=0] {	\includegraphics[trim={5.5cm 6cm  13.5cm 5cm},clip, width=1.3 \linewidth]{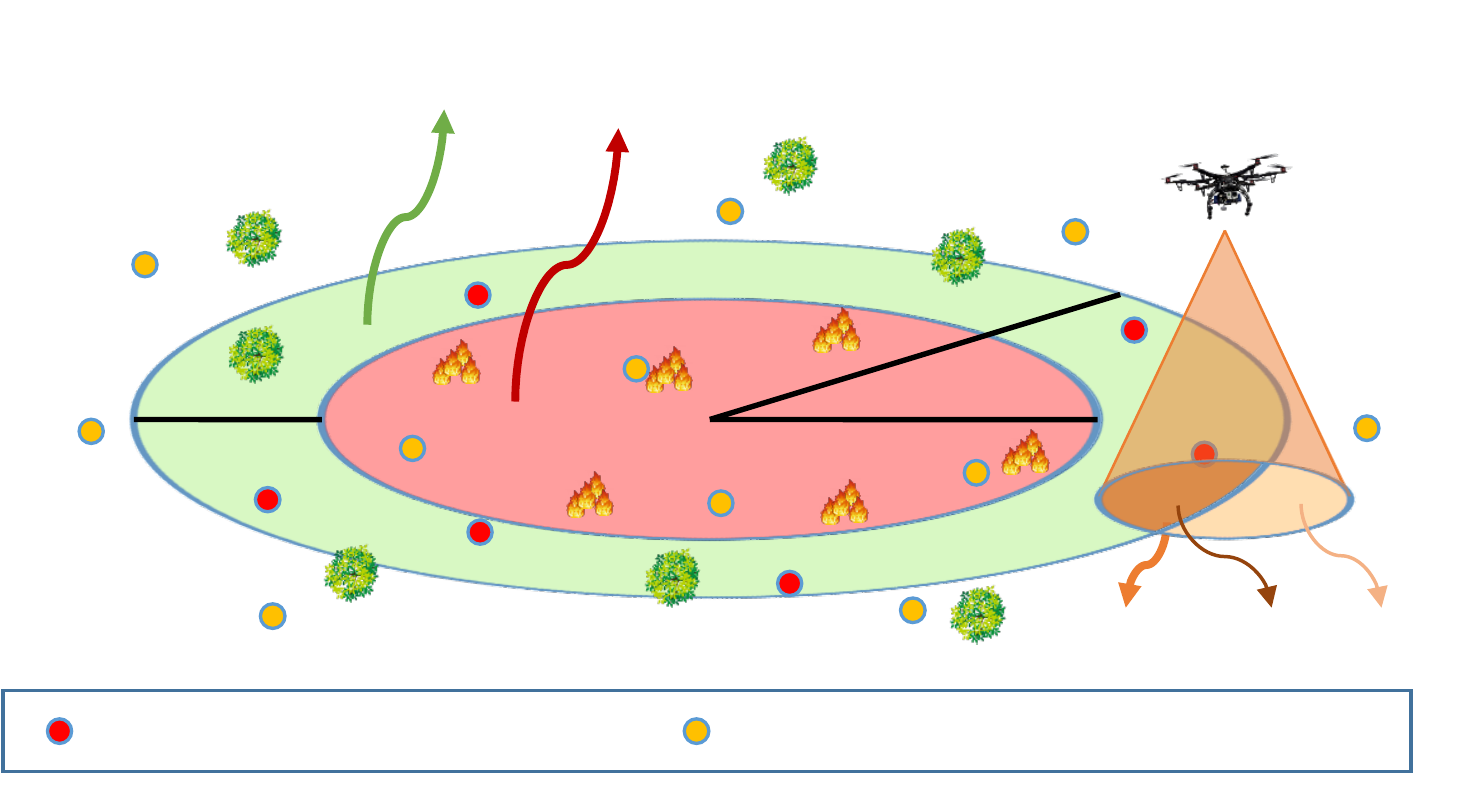}};
	\draw (-2.13,3.1) node {{$\mathcal{B}_s$}};
	\draw (-1.0,3.0) node {{$\mathcal{B}_f$}};
	\draw (4.3,-1.2) node {{$\mathcal{B}_{\rm hov}= \mathcal{B}_{\rm in} \cup \mathcal{B}_{\rm out}$}};	
	
	\draw (2.3,0.75) node {{$R_{f}$}};
	\draw (2.2,1.5) node {{$R_{s}$}};
	\draw (-3.8,0.75) node {{$d_s$}};
	
	\draw (2.63,-2.) node {{IoT devices not detecting fire}};
	\draw (-2.9,-2.) node {{IoT devices detecting fire}};		
	\end{tikzpicture}
	\caption{Illustration of the fire disk $\mathcal{B}_{f}$, the IoT sensor detection ring $\mathcal{B}_{s}$, and the UAV coverage region $\mathcal{B}_{\rm hov}$ which is divided into, $\mathcal{B}_{\rm in} =\mathcal{B}_{s} \cap \mathcal{B}_{\rm hov} $ and  $\mathcal{B}_{\rm out} =\mathcal{B}_{\rm hov} \setminus  \mathcal{B}_{\rm in}$. Fire detection at the UAV is only possible if $\mathcal{B}_{\rm in} \neq \emptyset$. } \label{fig:system_model}
\end{figure}

The IoT devices' locations are modeled according to a Poisson point process (PPP) distribution with density $\lambda_s$. For a forest of area $A$, the number of deployed IoT devices is $N_s = \lambda_s A$. The forest of interest is covered by $N_u$ UAVs such that each UAV searches over an area of $A/N_u$. Each UAV spends $T$ sec. hovering over one location to collect observations from sensors within its coverage region, denoted by $\mathcal{B}_{\rm hov}$, and moving to a new location. The UAV coverage region, $\mathcal{B}_{\rm hov}$ is a circular region centered at the UAV $x$-$y$ position with radius $R_{\rm hov}$ as shown in Fig.~\ref{fig:system_model}. The fire may be detected only if $\mathcal{B}_{\rm in} =\mathcal{B}_{s} \cap \mathcal{B}_{\rm hov} \neq  \emptyset $, which is the portion of the coverage area where there may be detecting sensors. The rest of the UAV coverage region is denoted as $\mathcal{B}_{\rm out} =\mathcal{B}_{\rm hov} \setminus  \mathcal{B}_{\rm in}$.

The design of the UAVs trajectory for fire detection is a complex optimization problem. To keep the analysis tractable at this early point of UAV-IoT network validation for wildfire detection, we assume random UAV locations such that UAVs are sufficiently far from each other. We also assume that the UAVs' locations at time step $k$ are independent of their locations at the previous time steps. Note that this is achievable in practice, and that performance can only improve if trajectory is optimized, which is consistent with our worst-case analysis. Next, we discuss the communication channel between the IoT devices and the UAVs. 

\subsection{IoT devices-UAV channel} 
\subsubsection{UAV height and coverage region} The UAV coverage region plays a significant role in the fire detection delay. While a larger $\mathcal{B}_{\rm hov}$ increases the probability that $\mathcal{B}_{\rm in} \neq \emptyset$, it also increases the required time $T$ to collect sensors' data within $\mathcal{B}_{\rm hov}$. Moreover, the sensors-UAV channel quality degrades as the coverage area increases. Let $P$ denote the IoT sensor transmission power and let $\sigma_n^2$ denote the receiver noise variance at the UAV, the average signal to noise ratio at the UAV is expressed as \cite{Akram2014Dec2}, 
\begin{align}\label{eq:SNR}
{\rm SNR} = \dfrac{P w^{-\alpha} }{\sigma_n^2} \left( \dfrac{p_{\rm LoS}}{\eta_{\rm LoS}} + \dfrac{1-p_{\rm LoS}}{\eta_{\rm NLoS}} \right),
\end{align}
where $w$ is the sensor-UAV distance, $\alpha$ is the path loss exponent, $\eta_{\rm LoS}$ ($\eta_{\rm NLoS}$) are the (non-)line of sight mean excessive path loss values, and $p_{\rm LoS}$ is the line of sight probability given by,
\begin{align}\label{eq:p_los}
p_{\rm LoS} = \dfrac{1}{1+a \exp\left( -b \left[\arcsin \left(\dfrac{h_{\rm hov} }{w}\right)-a \right] \right) } .
\end{align}
Here $h_{\rm hov}$ is the UAV height, and $a$ and $b$ are environmental parameters, for instance, $a= 4.88$ and $b= 0.43$ for suburban (forest) environment \cite{Akram2014Dec2}. Note that $p_{\rm LoS}$ increases as the distance between the transmitting IoT sensor and the UAV decreases for a fixed UAV height and environment parameters. From \eqref{eq:SNR} and \eqref{eq:p_los}, the lowest average SNR is at the edge of $\mathcal{B}_{\rm hov}$. In this work, for a given target SNR at the edge of $\mathcal{B}_{\rm hov}$, the UAV height is optimized to maximize the UAV coverage radius, $R_{\rm hov}$. Further, the transmission bit error rate $\epsilon_t$ is expressed as a function of the SNR for a given channel coding and  modulation scheme using \cite[p. 193]{Proakis2007}.

\begin{figure}
	\centering
	\begin{tikzpicture}[thick,scale=0.65, every node/.style={scale=0.65}]
	\draw (0, 0) node[inner sep=0] {	\includegraphics[width=1.5 \linewidth]{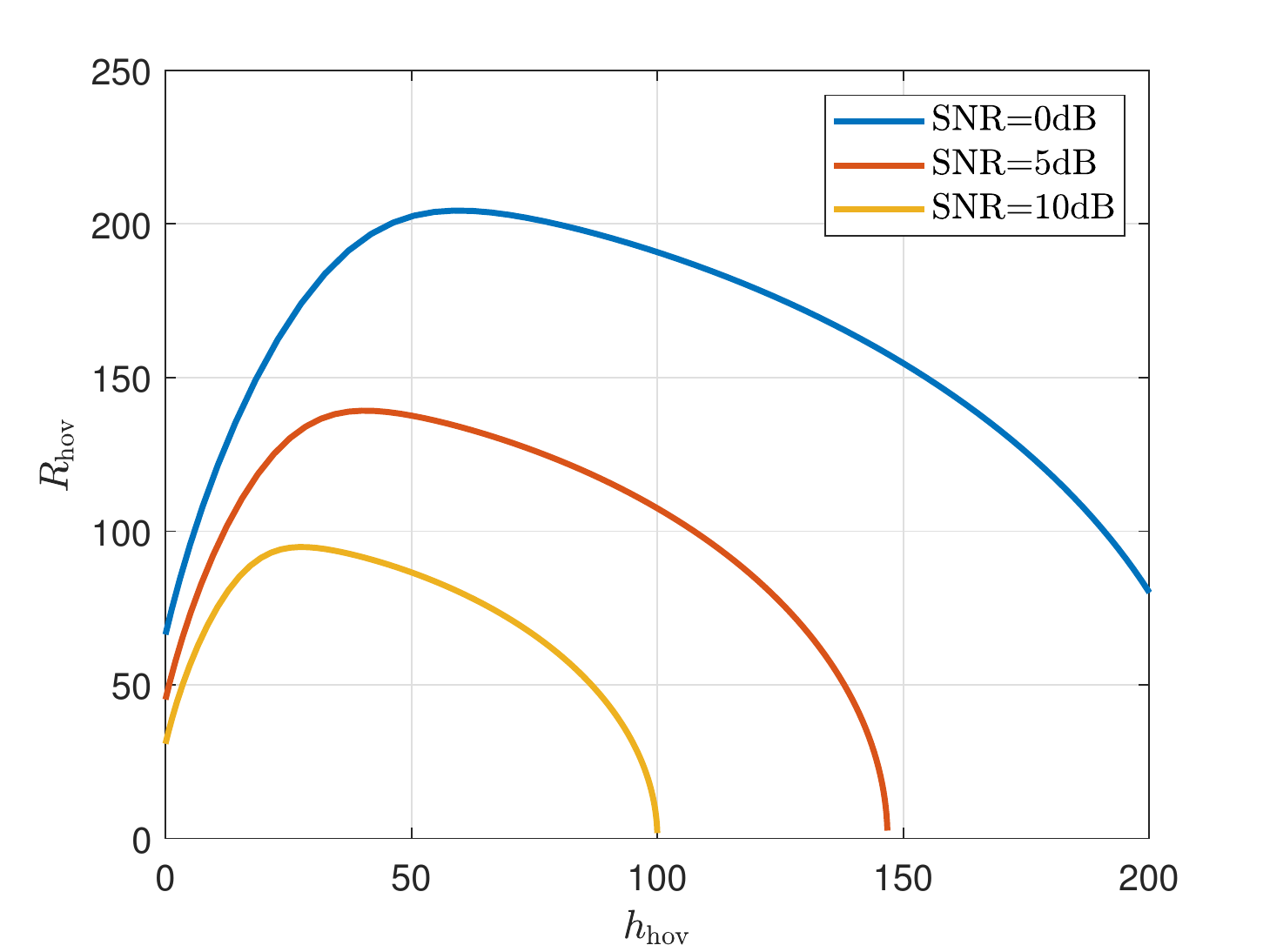}};
	\draw (-2.0,2.8) node {\huge{$\star$}};	
	\draw (-3.0,0.7) node {\huge{$\star$}};		
	\draw (-3.6,-.7) node {\huge{$\star$}};	
	\end{tikzpicture}
 \caption{Optimal UAV height for maximum area coverage for a target SNR at the edge of $\mathcal{B}_{\rm hov}$. Stars ($\star$) denote the optimal $h_{\rm hov}$ such that $R_{\rm hov}$ is maximized given the target SNR at the edge of the UAV coverage region.} \label{fig:optimal_uav_height}
\end{figure}
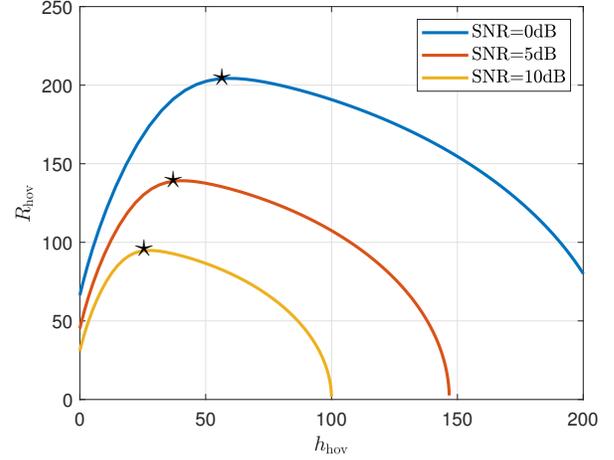

\underline{Example:} Let $P= 10$ dBm, $\sigma_n^2 = -90$ dB, $\eta_{\rm LoS}=0.1$ dB and $\eta_{\rm NLoS}=21$ dB. For target SNR values of $0$, $5$, and $10$ dB, the bit error rates, $\epsilon_{\rm BPSK}$, are $79\times10^{-2}$, $6\times10^{-3}$ and $3.9\times10^{-6}$, respectively, assuming BPSK modulation. Considering a repetition code with $\gamma$ denoting the number of bit repetitions such that $\dfrac{\gamma +1}{2} \in \mathbb{N}$, the transmission error is expressed as\footnote{Although more efficient modulation schemes and channel codes are available, BPSK modulation and repetition code are considered since we assume simple IoT device and since we we present a worst case fire detection study. },
\begin{align}
\epsilon_t = \sum_{i = \frac{\gamma +1}{2}}^\gamma {\gamma \choose i} \epsilon_{\rm BPSK}^{i}(1-\epsilon_{\rm BPSK})^{\gamma-i}.
\end{align} 
Further, for the target SNR values of $0$, $5$, and $10$ dB, the UAV height is optimized to maximize $R_{\rm hov}$ as shown in Fig.~\ref{fig:optimal_uav_height}.

The received signal at the UAV is subject to sensing and transmission errors with a total error,
\begin{align}
{\epsilon} = {\epsilon_s}(1-\epsilon_t)+ (1-{\epsilon_s}) \epsilon_t.
\label{eq:error}
\end{align}
While the sensing error $\epsilon_s$ is unavoidable, the transmission error $\epsilon_t$, can be decreased by reducing the $\mathcal{B}_{\rm hov}$ area. However, since all sensors deliver the same type of information, i.e., positive or negative fire sensing outcomes, and since sensors are self-powered and may have insufficient power for reliable transmission, a better engineering solution would be to allow certain probability of transmission error while increasing the coverage region to increase the number of collected measurements. At the UAV level, a positive decision (i.e., the UAV decides that a fire exists) is made if at least $M$ positive flags are received.

\subsection{Medium access control}
Slotted ALOHA is utilized as the medium access control to avoid excessive control overhead. The UAV first sends a wake-up/synchronization signal to the IoT devices under its coverage. Then, covered IoT devices transmit their observations with some transmission probability over a number of transmission slots before the UAV moves to the next hovering location. 

The UAV spends $T = T_{\rm hover} + T_{\rm travel}$ sec. to collect data from one coverage region and then travel to the next hovering location, where $T_{\rm hover}$ and $ T_{\rm travel}$ are the average hovering and traveling times, respectively. During $T_{\rm hover}$ sec., the UAV collects an average of $N= \beta \lambda_s \pi R_{\rm hov}^2$ IoT sensor observations from its coverage region where $\beta \in [0,1]$ is a design parameter representing the ratio of the number of collected observations to the total number of covered sensors. The hovering time is expressed as,
\begin{align}
T_{\rm hover} = \gamma  \eta  T_{\rm sym} N
\end{align} 
where $\eta=36.8 \% $ is the efficiency of slotted Aloha, $T_{\rm sym}$ is the time required to transmit one symbol and $\gamma$ is the number of bit repetitions assuming repetition code. For convenience, we denote $ T_{\rm obs} = \gamma  \eta  T_{\rm sym} $ as the time needed to collect one observation. The hovering time $T_{\rm hover}$ is designed such that $\beta \approx 1$.\footnote{Reducing $\beta$ has the same effect as reducing $\lambda_s$. However, a low $\lambda_s$ is more practical as it reduces the system cost.} Therefore, as $\lambda_s$ increases, $T_{\rm hover}$ also increases to allow enough time to collect the same ratio of observations within $\mathcal{B}_{\rm hov}$. Next, we mathematically formulate the problem statements.

\section{Problem statements}\label{sec:problem_stat}
 
Now we are ready to define two optimization problems which improve the system performance. The first one seeks to maximize the fire detection probability within a limited time given a predefined budget. The second problem seeks to minimize the overall losses caused by a possible wildfire. These losses include the damage the fire causes, the fire fighting cost,  and the system cost. The two optimization problems are defined mathematically as follows:
\subsubsection{Wildfire detection probability maximization} Given a limited system budget for sensors and UAVs installation (and maintenance), the objective is to maximize the fire detection probability within a limited time frame, $T_f$. Practically, the time can be determined by fire fighting departments, based on the critical time beyond which  putting down the fire becomes too costly, or it becomes out of control. This problem can be mathematically expressed as,
	\begin{align}
	 {\textbf P1:} \quad & \underset{M,N_s N_{u}}{\max}  \ \quad {\pi}_D[K] ,  \nonumber \\
	 & \quad s.t. \quad  \quad \omega_s N_s + \omega_u N_{u} \leq \zeta, \label{eq:cost} 
	\end{align}
	where ${\pi}_D[K] $ is the probability of fire detection by time step $K$ after the fire ignition (i.e., at any time step $k\leq K$) with $K= \left\lfloor \dfrac{T_f}{T} \right\rfloor$. The optimization in {\textbf{P1}} is with respect to the number $M$ of fire flags required to make a positive decision at the UAV, the number $N_s$ of sensors, and the number $N_u$ of UAVs. The constraint \eqref{eq:cost} restrict the system cost to $\zeta$ where $\omega_s$ and $\omega_u$ are the sensor and the UAV costs.
\subsubsection{Wildfire losses minimization} The objective of this optimization problem is to minimize the total losses caused by a probable wildfire. These losses include the fire damage (which may include damage to land and property), the cost of fire fighting, and the UAV-IoT system costs. The cost of fire damage and fire fighting, denoted as $\omega_D[k]$, is a monotonically increasing function with time, since damage to land and property increases with the fire area which increases with time, and also the cost of fire fighting increases with the fire area. We assume that there is a maximum time $T_D$ after which the fire is detected by other methods such as satellite imaging. Therefore, there is no additional cost at any time after $T_D$. This problem is mathematically written as,
	\begin{align}\label{eq:p2}
	{\textbf P2:} \quad \underset{M,N_s N_{u}}{\min}   \quad &\omega_s N_s + \omega_{u} N_{u}  + \left( \sum_{k=1}^{\bar{K}} \omega_D[k]  {\rho}_D[k] \right) \nonumber \\
	& + \omega_D[\bar{K}+1] \left(1 -\pi_D[\bar{K}] \right) ,
	\end{align}
	where ${\rho}_D[k] $ is the probability of fire detection exactly at the time step $k$, and $\bar{ K} = \left\lfloor \dfrac{T_D}{T} \right\rfloor$ represents the number of time steps before the fire is detected by other methods. In \eqref{eq:p2}, the first two terms describe the UAV-IoT network cost, the third term describes the fire detection costs within the period $T_D$, and the last term represents the cost of not detecting the fire by $T_D$. 

In the next section, we derive mathematical expressions for $\pi_D[K]$ and $\rho_D[k]$. Then, the wildfire detection probability maximization and the wildfire losses minimization problems are solved by performing simple search algorithms.

\section{Detection Performance Analysis}\label{sec:performance_analysis}

In this section, we derive the wildfire detection probabilities $\rho_D[k]$ and $\pi_D[k]$ in terms of the number of positive flags needed to declare fire detection at the UAV, $M$, and the number of IoT devices $N_s$ and UAVs $N_u$. This is done in four steps as presented in the next four subsections. 

\subsection{Intersection between UAV coverage region and IoT sensors detection ring:}

In order to analyze the probability of fire detection at the UAV level, we first evaluate the probability of intersection between the UAV coverage region and the IoT sensors detection ring $\mathcal{B}_{\rm in}[k] = \mathcal{B}_s[k] \cap \mathcal{B}_{\rm hov}[k] \neq \emptyset$. For a circular fire with radius $R_{f}[k]$, any sensor located in the sensor detection ring, $\mathcal{B}_s[k]$, with inner and outer radii of $R_{f}[k]$ and $R_{s}[k] = R_{f}[k]+d_s$, respectively, detects the fire with probability $1- \epsilon_s$. The fire can be detected at the UAV only if $\mathcal{B}_{\rm in}[k] = \mathcal{B}_s[k] \cap \mathcal{B}_{\rm hov}[k] \neq \emptyset$. We define the region within which the UAV coverage intersects with the IoT sensors ring as follows. 

\begin{figure}
	\centering
	\begin{tikzpicture}[thick,scale=0.7, every node/.style={scale=0.7}]
	\draw (0, .7) node[inner sep=0] {	\includegraphics[trim={5.5cm 6cm  13.5cm 5cm},clip, width=1.33 \linewidth]{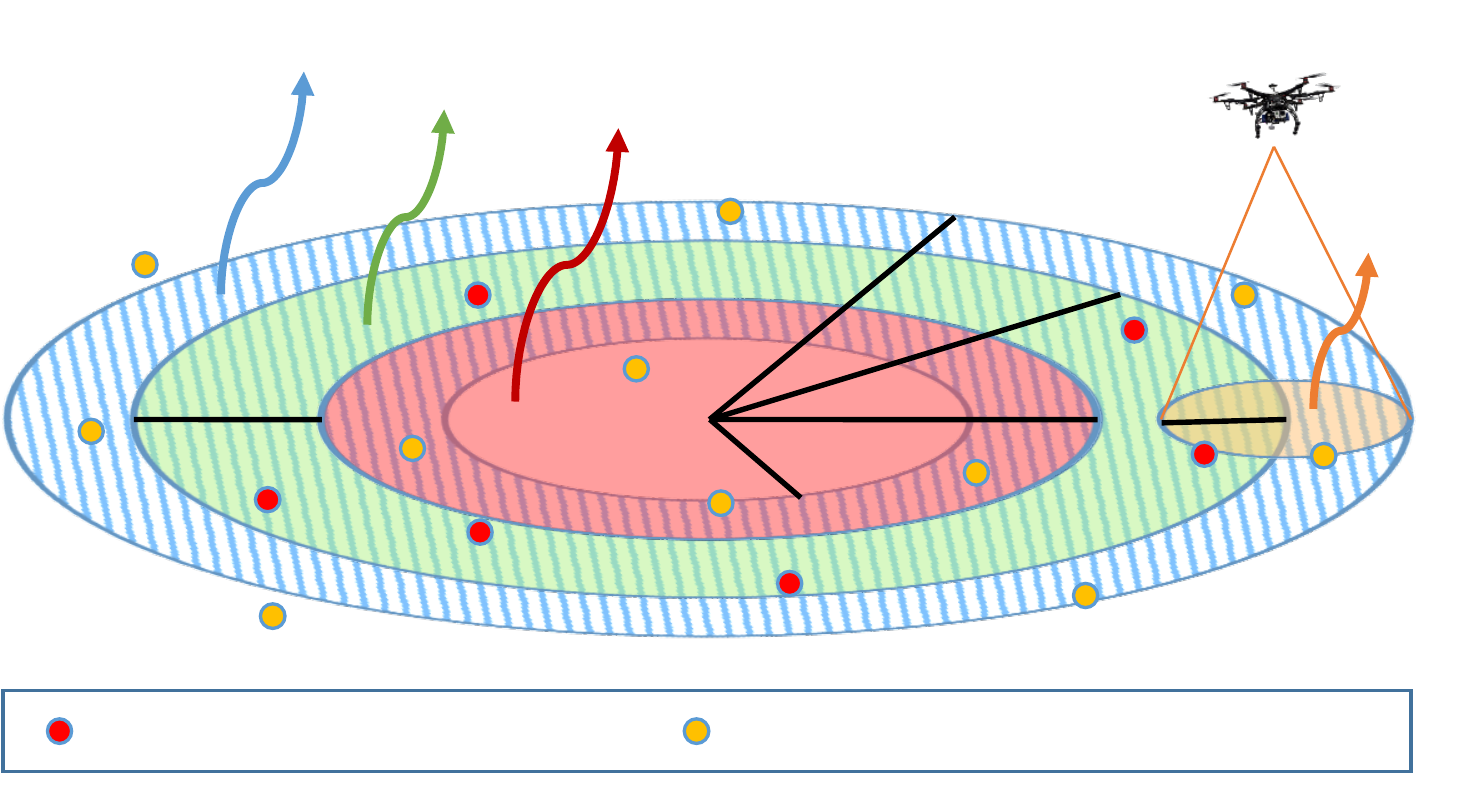}};
	\draw (-3.3,3.4) node {{$\mathcal{B}_u$}};
	\draw (-2.13,3.1) node {{$\mathcal{B}_s$}};
	\draw (-1.0,3.0) node {{$\mathcal{B}_f$}};
	\draw (5.13,2.1) node {{$\mathcal{B}_{\rm hov}$}};	
	
	\draw (0.8,0.2) node {{$\underline{R}_{u}[k]$}};
	\draw (2.3,0.75) node {{$R_{f}[k]$}};
	\draw (2.2,1.5) node {{$R_{s}[k]$}};
	\draw (0.7,1.8) node {{$\overline{R}_{u}[k]$}};
	\draw (4.1,0.75) node {{$R_{\rm hov}$}};
	\draw (-3.8,0.75) node {{$d_s$}};
	
	\draw (2.63,-2.) node {{IoT devices not detecting fire}};
	\draw (-2.9,-2.) node {{IoT devices detecting fire}};	
	\end{tikzpicture}
	\caption{The figure illustrates of the UAV detection ring $\mathcal{B}_u$ shaded in blue. The coverage region of a UAV $\mathcal{B}_{\rm hov}$ intersects with sensor detection ring $\mathcal{B}_s$ if the UAV $x-y$ location belongs to $\mathcal{B}_u$.} \label{detection_rings}
\end{figure}
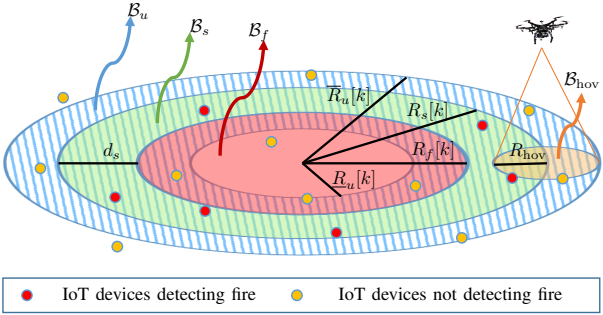

{\definition The UAV detection ring at time $k$ is the set of UAV locations within which the UAV coverage region intersects with the sensors detection ring. The UAV detection ring is defined as,
	\begin{align}
	\mathcal{B}_u[k] = \{ (r,\theta): r \in \left[\underline{R}_{u}[k],\overline{R}_{u}[k] \right], \quad \theta \in [0,2\pi]  \},
	\end{align}	
	where $\underline{R}_{u}[k] = \min(0,R_{f}[k]-R_{\rm hov} )$ and $ \overline{R}_{u}[k] = R_{s}[k]+R_{\rm hov}$. The UAV detection ring is shown in Fig.~\ref{detection_rings}.  \label{def:B_u} }

The probability that the UAV is located over the ring $\mathcal{B}_u[k]$ is given as follows,
{\proposition
Given that each UAV is at a uniformly random location within its coverage portion at any time step $k$, the probability of $\mathcal{B}_{\rm in}[k] \neq  \emptyset $ is,
\begin{align}\label{eq:p_u}
P_{\rm int}[k] = \dfrac{N_{u} A_u[k]}{A} ,
\end{align}
where ${A}_{u}[k] = \pi (\overline{R}_{u}^2[k] - \underline{R}_{u}^2[k])$, as defined in Definition~\ref{def:B_u}, is the area of $\mathcal{B}_u[k]$ and $A$ is the total forest area. }

Since error may occur in the data sensing and transmission, a false fire alarm can be indicated at the UAV while $\mathcal{B}_{\rm in}[k] = \emptyset$ if at least $M$ false positive flags are received at the UAV. If, on the other hand, the UAV indicates a fire alarm while $\mathcal{B}_{\rm in}[k] \neq \emptyset$, we consider that the fire is correctly detected whether the fire alarm was influenced by the sensors within $\mathcal{B}_{\rm in}[k]$ or the faulty sensors within $\mathcal{B}_{\rm out}[k]$. Once the UAV indicate a fire alarm by receiving at least $M$ positive flags, the system enters a verification mode where a robust decision is made at the UAV. This is discussed in the following subsection.

\subsection{Markov Representation}
\begin{figure}
	\centering
	\begin{tikzpicture}[thick,scale=.9, every node/.style={scale=.9}]
	\draw (0, 0) node[inner sep=0] {\includegraphics[trim={4cm 7cm  5cm 4cm},clip, width=.95\linewidth]{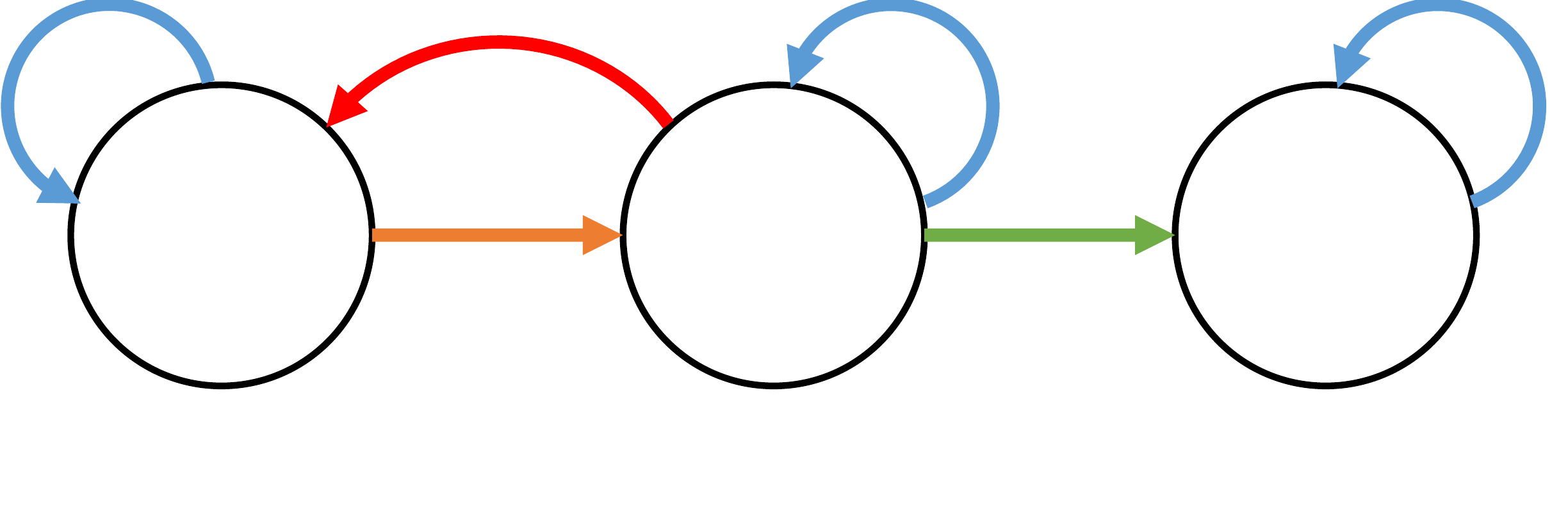}};
	\draw (-3,0.2) node {{$\mathcal{N}$}};		
	\draw (-.1,0.2) node {{$\mathcal{V}$}};	
	\draw (2.8,0.2) node {{$\mathcal{D}$}};	
	
	\draw(3.3,1.5)node{\color{NavyBlue}{\small $1$}};
	\draw(0.3,1.5)node{\color{NavyBlue}{\small $P_{VV}[k]$}};
	\draw(-3.3,1.5)node{\color{NavyBlue}{\small $P_{NN}[k]$}};
	\draw (-1.5,1.4)node{\color{Red}{\small $P_{VN}[k]$}};
	\draw(-1.5,-0.2)node{\color{Orange}{\small $P_{NV}[k]$}};
	\draw(1.4,-.2) node {\color{LimeGreen}{\small $P_{VD}[k]$}};	
	\end{tikzpicture}
	\caption{Markov representation of the the wildfire detection model. Note that the transition probabilities are time dependent.} \label{fig:DTMC}
\end{figure}
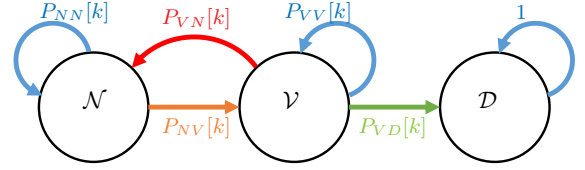

The state of fire detection can be represented by a time-inhomogeneous discrete time Markov chain (DTMC) with the state space $\mathcal{S} = \{\mathcal{N}, \mathcal{V}, \mathcal{D} \}$, where $\mathcal{N}, \mathcal{V}$ and $ \mathcal{D}$ represent no fire, verification, and fire detection states, respectively. The DTMC is time-inhomogeneous because the transition probabilities are time dependent. Note that any time-inhomogeneous DTMC can be converted to time-homogeneous DTMC by extending the state space over time. The wildfire detection model is illustrated in Fig.~\ref{fig:DTMC}.

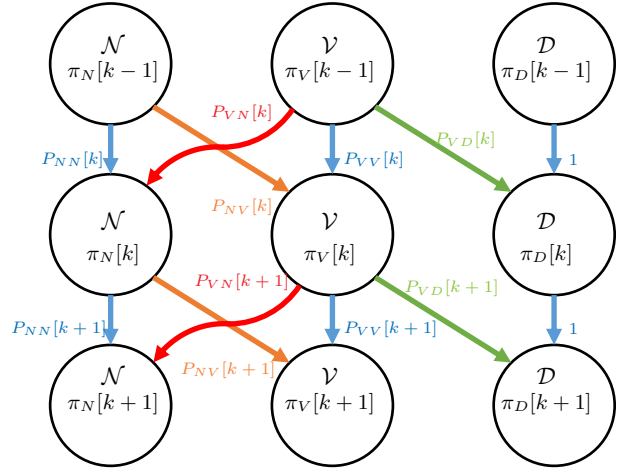
\begin{figure}
	\centering
	\begin{tikzpicture}[thick,scale=.9, every node/.style={scale=.9}]
	\draw (0, 0) node[inner sep=0] {\includegraphics[trim={4cm 0cm  5cm 0cm},clip, width=1.05\linewidth]{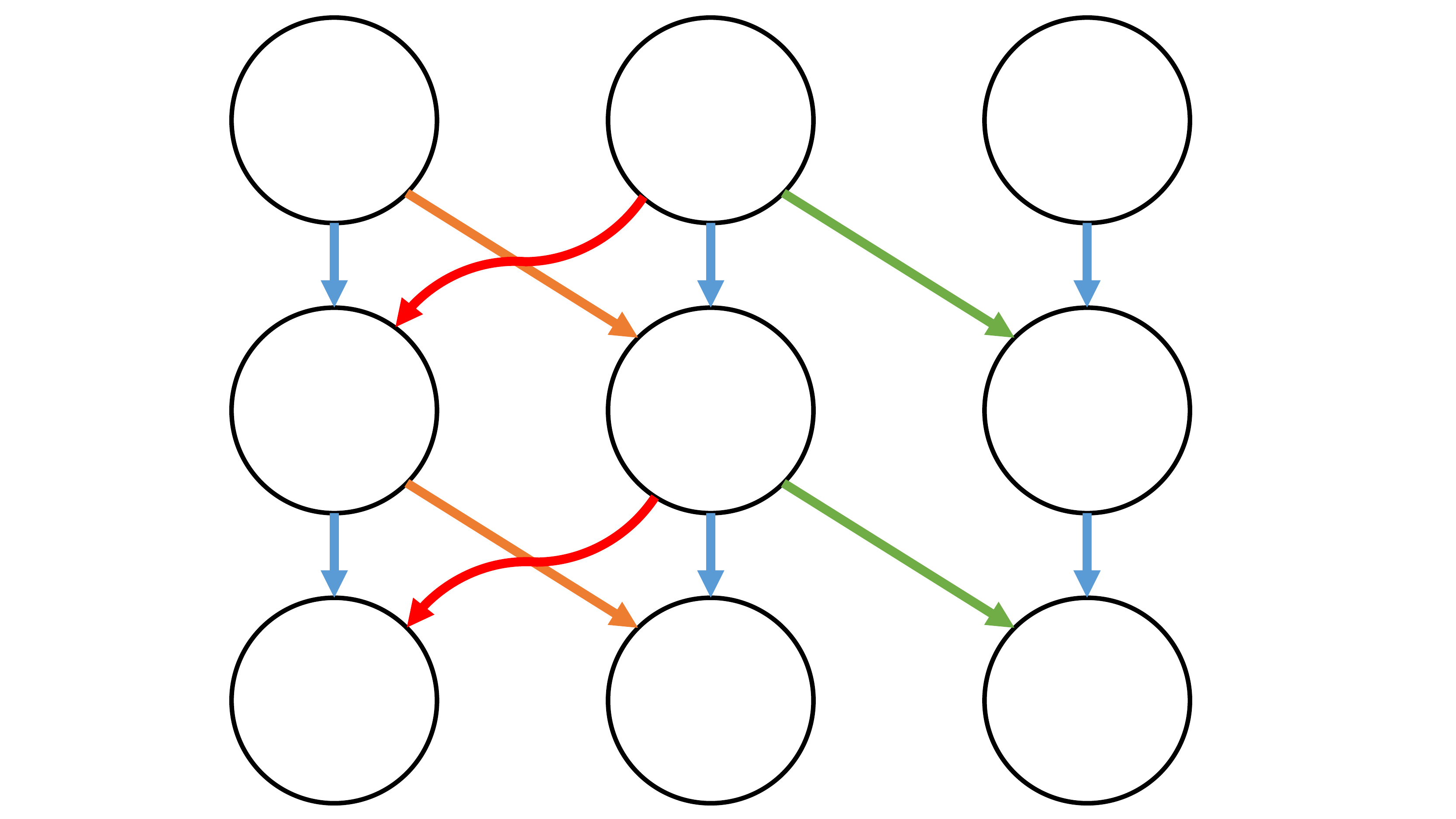}};
	\draw (-3.2,2.8) node {{$\mathcal{N}$}};		
	\draw (-.0,2.8) node {{$\mathcal{V}$}};	
	\draw (3.2,2.8) node {{$\mathcal{D}$}};	
	\draw (-.0,2.4) node {\small $\pi_V[k-1]$};
	\draw (-3.2,2.4) node {\small $\pi_N[k-1]$};
	\draw (3.2,2.4) node {\small $\pi_D[k-1]$};
	
	\draw (-3.2,0.2) node {{$\mathcal{N}$}};		
	\draw (-.0,0.2) node {{$\mathcal{V}$}};	
	\draw (3.2,0.2) node {{$\mathcal{D}$}};	
	\draw (-.0,-0.3) node {\small $\pi_V[k]$};
	\draw (-3.2,-0.3) node {\small $\pi_N[k]$};
	\draw (3.2,-0.3) node {\small $\pi_D[k]$};
	
	\draw (-3.2,-2.1) node {{$\mathcal{N}$}};		
	\draw (-.0,-2.1) node {{$\mathcal{V}$}};	
	\draw (3.2,-2.1) node {{$\mathcal{D}$}};	
	\draw (-.0,-2.5) node {\small{$\pi_V[k+1]$}};
	\draw (-3.2,-2.5) node {\small{$\pi_N[k+1]$}};
	\draw (3.2,-2.5) node {\small{$\pi_D[k+1]$}};
	
	\draw(3.6,1.1)node{\color{NavyBlue}{\scriptsize  $1$}};
	\draw(0.65,1.1)node{\color{NavyBlue}{\scriptsize  $P_{VV}[k]$}};
	\draw(-3.8,1.1)node{\color{NavyBlue}{\scriptsize  $P_{NN}[k]$}};
	\draw (-1.3,1.8)node{\color{Red}{\scriptsize  $P_{VN}[k]$}};
	\draw(-1.3,0.4)node{\color{Orange}{\scriptsize  $P_{NV}[k]$}};
	\draw(2,1.4) node {\color{LimeGreen}{\scriptsize  $P_{VD}[k]$}};
	
	\draw(3.6,1.1-2.5)node{\color{NavyBlue}{\scriptsize $1$}};
	\draw(0.9,1.1-2.5)node{\color{NavyBlue}{\scriptsize $P_{VV}[k+1]$}};
	\draw(-4,1.1-2.5)node{\color{NavyBlue}{\scriptsize $P_{NN}[k+1]$}};
	\draw (-1.3,1.8-2.5)node{\color{Red}{\scriptsize $P_{VN}[k+1]$}};
	\draw(-1.5,0.4-2.4)node{\color{Orange}{\scriptsize $P_{NV}[k+1]$}};
	\draw(1.8,1.4-2.2) node {\color{LimeGreen}{\scriptsize $P_{VD}[k+1]$}};	
	\end{tikzpicture}
	\caption{State probability at arbitrary time.} \label{fig:time_inhomo_DTMC_state_prob}
\end{figure}

Denote the probability of being at the states $\{\mathcal{N}, \mathcal{V}, \mathcal{D}\} $ at time step $k$ by the vector $\boldsymbol{\pi}[k] = [\pi_N[k] \, \pi_V[k]\, \pi_D[k]]$ with $\boldsymbol{\pi}[0] = [1 \, 0 \, 0]$, i.e., the "no fire" state is assumed initially. Based on the model in Fig.~\ref{fig:DTMC}, the transition matrix is expressed as, 

\begin{align}
{\bf{P}}[k] = 
\begin{bmatrix}\label{eq:transition_matrix}
P_{NN}[k] & P_{NV}[k] & 0 \\
P_{VN}[k] & P_{VV}[k] & P_{VD}[k] \\
0 & 0 & 1
\end{bmatrix},
\end{align}
where $P_{i,j}[k]$ is the probability of transition from state $i$ to state $j$ at time step $k$. The transition probability at arbitrary time is described in Fig.~\ref{fig:time_inhomo_DTMC_state_prob}.

The probability of being at any state at time step $K$ is given by,
\begin{align}
\boldsymbol{\pi}[K] = \boldsymbol{\pi}[0] {\bf{P}}[1] {\bf{P}}[2] \cdots {\bf{P}}[K].
\end{align}
Since the detection state is an absorbing state, the probability of fire detection by the time step $K$ is ${\pi}_D[K] $. The probability of fire detection exactly at time step $K$ is given by,
\begin{align} 
\rho_D[K] &= \pi_D[K] - \pi_D[K-1] \label{eq:detection_at_k_1},  \\
&= {\pi}_V[K-1] P_{VD}[K]. \label{eq:detection_at_k_2}
\end{align}
where \eqref{eq:detection_at_k_1} follows for the fact that state $\mathcal{D}$ is an absorbing state and \eqref{eq:detection_at_k_2} is obtained using the fact that state $\mathcal{D}$ is reached for the first time at time $K$ from the state $\mathcal{V}$ at time $K-1$.
 
To calculate $\pi_D[K]$ and $\rho_D[K]$, we need to derive the transition probabilities in \eqref{eq:transition_matrix} to solve the optimization problems discussed in the previous section. The system stays in state $\mathcal{N}$ until a fire is detected or a false alarm is indicated. Therefore, the transition probabilities $P_{NV}[k]$ and $P_{NN}[k]$ are expressed as,
\begin{align}
P_{NV}[k] &= P_{\rm d}[k] + P_{\rm fa}[k], \label{eq:P_NV}\\
P_{NN}[k] &= 1 - P_{NV}[k], \label{eq:P_NN}
\end{align}
where $ P_{\rm d}[k] $ and $ P_{\rm fa}[k] $ are the detection and false alarm probabilities at time step $k \in \{1, \cdots, K\}$.

Once in the verification state, the UAV spends an average time of $T_{\rm vrf} \geq T$ until a robust decision regarding the fire detection is made. Based on $T_{\rm vrf} $, the transition probabilities $P_{VV}[k]$, $P_{VF}[k]$ and $P_{VD}[k]$ are modeled as, 
\begin{align}
P_{VV}[k] &= 1 - \dfrac{T}{T_{\rm vrf}}, \label{eq:P_VV}\\ 
P_{VN}[k] &= (1-P_{VV}[k]) \dfrac{P_{\rm fa}[k]}{P_{\rm d}[k]+P_{\rm fa}[k]} , \label{eq:P_VN}\\
P_{VD}[k] &= (1-P_{VV}[k]) \dfrac{P_{\rm d}[k]}{P_{\rm d}[k]+P_{\rm fa}[k]}. \label{eq:P_VD}
\end{align}

In the next subsection, the detection and false alarm probabilities $P_{\rm d}[k]$ and $P_{\rm fa}[k]$ are derived.

\subsection{Detection and False Alarm Probabilities}
As the UAV collects $N$ observation per hovering location, it recognizes a fire possibility if there are at least $M$ positive flags. The optimization of $M$ is essential. As $M$ increases, both the false alarm and the fire detection probabilities decrease. The false alarm probability is derived as follows.

\subsubsection{False Alarm probability $P_{\rm fa}[k]$}
A false alarm is possible only if $\mathcal{B}_{\rm in}[k]=\emptyset $. Otherwise, any fire detection at the UAV is considered as a successful fire detection. A false alarm occurs when the UAV receives at least $M$ false positive flags although $\mathcal{B}_{\rm in}[k]=\emptyset $. These faulty positive flags are caused by the sensing and transmission errors, $\epsilon$.
{\proposition Assuming independent sensing and transmission errors, $\epsilon$ defined in \eqref{eq:error}, among the transmitting $N$ sensors within $\mathcal{B}_{\rm hov}$, the probability of receiving at least $M$ positive flags given $\mathcal{B}_{\rm in}=\emptyset $ is given by the binomial expression,
\begin{align}\label{eq:P_fa}
P_{{\rm fa}|{\rm \overline{int} } }= \sum_{m=M}^{N} {N \choose m} \epsilon^{m}(1-\epsilon)^{N-m}.
\end{align}
}

From \eqref{eq:p_u} and \eqref{eq:P_fa}, the probability of false alarm at the UAV is given by,
\begin{align}
P_{\rm fa}[k]  &= \left(1-P_{\rm int}[k] \right) \sum_{m=M}^{{N}} {{N} \choose m} \epsilon^{m}(1-\epsilon)^{{N}-m}, \label{eq:Binomial0} \\
&= \left(1-P_{\rm int}[k] \right) \left(1-\sum_{m=0}^{M-1} {{N} \choose m} \epsilon^{m}(1-\epsilon)^{{N}-m} \right). \label{eq:Binomial}
\end{align}
The two expressions in \eqref{eq:Binomial0} and \eqref{eq:Binomial} are equivalent. We use \eqref{eq:Binomial0} or \eqref{eq:Binomial} depending on whether $M>N/2$ or not.

\subsubsection{Detection Probability $P_{\rm d}[k]$}
After each time interval of $T$ sec., the size of the fire grows according to the fire spreading model. The fire can be detected only if $\mathcal{B}_{\rm in}[k] \neq \emptyset$. The probability $P_{\rm d}[k]$ is therefore expressed as, 
\begin{align}\label{eq:P_uav} 
P_{\rm d}[k] = P_{\rm int}[k] P_{{\rm d}|{\rm int}}[k],
\end{align}
where $P_{{\rm d}|{\rm int}}[k]$ is the conditional probability of fire detection given $\mathcal{B}_{\rm in} \neq \emptyset$ which depends on the average number of IoT sensors within $\mathcal{B}_{\rm in}$ (denoted as $n_{\rm in}$). Therefore, $P_{{\rm d}|{\rm int}}[k]$ can be expressed as
\begin{align}\label{eq:P_d_given_B_in}
 P_{{\rm d}|{\rm int} }[k] = \sum_{n_{in}=0}^{N} {P}_{n_{\rm in}|{\rm int} }[k] P_{ {\rm d}|n_{\rm in} },
\end{align}
where ${P}_{n_{\rm in}|{\rm int} }[k] $ is the probability of having an average of $n_{\rm in}$ IoT sensors inside $\mathcal{B}_{\rm in}$ and $P_{ {\rm d}|n_{\rm in} }$ is the probability of collecting at least $M$ positive flags given that $n_{\rm in}$ sensors are located in $\mathcal{B}_{\rm in}$.

{\proposition For independent sensing and transmission errors, $\epsilon$, among the transmitting $n_{\rm in}$ IoT sensors within $\mathcal{B}_{\rm in}$ and $n_{\rm out} =N-n_{\rm in}$ IoT sensors within $\mathcal{B}_{\rm out}$, the probability of receiving at least $M$ positive flags $P_{ {\rm d}|n_{\rm in} }$ is equal to the probability of receiving $m_{\rm in}$ positive flags from $\mathcal{B}_{\rm in}$ multiplied by the probability of receiving $m_{\rm out}$ positive flags from $\mathcal{B}_{\rm out}$ such that $m_{\rm in} + m_{\rm out} \geq M$. This probability is expressed based on the Poisson binomial distribution as\footnote{Poisson binomial distribution describes the number of successes out of $n$ independent trials where each event have a probability of success $P_i, \, \forall i\in {1, \cdots, n}$. In our model, a group of $n_{\rm in}$ sensors have a success rate $(1-\epsilon)$ while ${N}-n_{\rm in}$ sensors have success rate of $\epsilon$.},
	{\small
\begin{align}
P_{ {\rm d}|n_{\rm in} } = & \sum_{m_{\rm in}=0}^{n_{\rm in}} {n_{\rm in} \choose m_{\rm in}} \epsilon^{n_{\rm in}-m_{\rm in}}(1-\epsilon)^{m_{\rm in}}  \nonumber \\ 
& \sum_{m_{\rm out}=M-m_{\rm in}}^{n_{\rm out}} {n_{\rm out} \choose m_{\rm out}} \epsilon^{m_{\rm out}}(1-\epsilon)^{n_{\rm out}-m_{\rm out}}, \\
=& \; 1- \sum_{m_{\rm in}=0}^{M-1} {n_{\rm in} \choose m_{\rm in}} \epsilon^{n_{\rm in}-m_{\rm in}}(1-\epsilon)^{m_{\rm in}} \nonumber \\ 
& \sum_{m_{\rm out}=0}^{M-m_{\rm in}-1} {n_{\rm out} \choose m_{\rm out}} \epsilon^{m_{\rm out}}(1-\epsilon)^{n_{\rm out}-m_{\rm out}} \label{eq:Poisson_Binomial_2}, 
\end{align} }
where $m_{\rm in}$ is the number of true positive flags and $m_{\rm out}$ denotes the number of false positive flags collected at the UAV from $\mathcal{B}_{\rm in}$ and $\mathcal{B}_{\rm out}$, respectively. For $M<N/2$, it is more computationally efficient to use \eqref{eq:Poisson_Binomial_2} to calculate $P_{ {\rm d}|n_{\rm in} }$.  }

It remains to calculate ${P}_{n_{\rm in}|{\rm int} }[k] $ which is derived in the next subsection. 

\subsection{The probability of having an average of $n_{\rm in}$ sensors inside $\mathcal{B}_{\rm in}$, ${P}_{n_{\rm in}|{\rm int} }[k] $ }
Since the IoT devices are PPP distributed, the average number of IoT devices located inside $\mathcal{B}_{\rm in}$ is $n_{\rm in}= \lambda_s A_{\rm in}$, where $A_{\rm in}$ is the area of $\mathcal{B}_{\rm in}$. Since $n_{\rm in}$ is discrete, we can express the probability that $n_{\rm in}=  \lfloor \lambda_s A_{\rm in} \rfloor $ as
\begin{align}\label{eq:A_in_bounds}
{P}_{n_{\rm in}|{\rm int} }[k]  = \mathbb{P}( n_{\rm in}/ \lambda_s \leq A_{\rm in}[k] \leq (n_{\rm in} +1)/ \lambda_s).
\end{align} 
To proceed, we need to find the probability density function (PDF) of $A_{\rm in}[k]$. We note that $A_{\rm in}[k]$ is the area of $\mathcal{B}_{rm in}$ and so a function of the distance between the fire center and the $x$-$y$ position of the UAV, denoted as $R$. We drop the time index for convenience and explicitly express $A_{\rm in}(R)$ through its dependence on $R$. As the fire center and the UAV locations at any time step are uniformly distributed, the PDF of $R$ given that $\mathcal{B}_{\rm in} \neq \emptyset$ is expressed as,
\begin{align}\label{eq:R_pdf}
f_R(r|\underline{R}_{u}[k] \leq r \leq \overline{R}_{u}[k]) = \dfrac{2r}{\overline{R}_{u}^2[k]-\underline{R}_{u}^2[k]}.
\end{align}
The area ${A}_{\rm in}$ is expressed as a function of $R$ as,
\begin{align}\label{eq:A_in}
{A}_{\rm in}(R) = {C}_{\rm int}(R_s, R_{\rm hov};R) -{C}_{\rm int}(R_f, R_{\rm hov};R),
\end{align}
where ${C}_{\rm int}(R_i, R_{\rm hov};R)$ is the area of intersection between two circles with radii $R_i$ and $R_{\rm hov}$ whose centers are separated by a distance $R$.
{\definition The area of intersection between two circles with radii $R_i$ and $R_{\rm hov}$ and centers separated by distance $R$ is expressed as \cite{Bushnaq2019, Flint2017},
\begin{align}\label{eq:c_int}
{C}_{\rm int}(R_i, R_{\rm hov};R) = \int_{0}^{R_{i}} r \theta({r}) dr,
\end{align}
where,
{\small
\begin{align}\label{eq:theta}
&\theta(r) = \nonumber \\
&\left\{ \begin{array}{cc} 
2\pi & 0\leq r \leq \max(0,R_{\rm hov}-R), \\
2\arccos\left(\dfrac{R^2+r^2-R_{\rm hov}^2}{2R r}\right) & |R_{\rm hov}-R| \leq r \leq R_{\rm hov}+R ,\\
0 & \text{otherwise}. 
\end{array} \right. 
\end{align} \label{def:circle_intersection} } }
Based on Definition \ref{def:circle_intersection}, $A_{\rm in}(R)$ is expressed as, 
\begin{align}
A_{\rm in}(R) = \int_{0}^{R_{s}} r \theta({r}) dr - \int_{0}^{R_{f}} r \theta({r}) dr.
\end{align}

By combining \eqref{eq:A_in_bounds} and \eqref{eq:R_pdf},  ${P}_{n_{\rm in}|{\rm int} }[k] $ is expressed as, 
\begin{align}
{P}_{n_{\rm in}|{\rm int} }[k] &= \mathbb{P}( n_{\rm in}/ \lambda_s \leq A_{\rm in}[k] \leq (n_{\rm in} +1)/ \lambda_s), \nonumber \\ \label{eq:P_N_in_numerical}
=&  \int_{r\in \, \left\{ R: \, A_{\rm in}(R) \in \, \left[\dfrac{n_{\rm in}}{\lambda_s}, \dfrac{n_{\rm in}+1}{\lambda_s}\right] \right\} } \dfrac{2r}{\overline{R}_{u}^2[k]-\underline{R}_{u}^2[k]} dr.
\end{align}
Given the complex relation between $R$ and $A_{\rm in}$, it is difficult to express $R$ as a function of $A_{\rm in}$ in a closed form expression or to express the PDF of $A_{\rm in}$ through a change of variables. Alternatively, we can calculate the probability ${P}_{n_{\rm in}|{\rm int} }[k]$ numerically by solving \eqref{eq:A_in} to find the range of $R$ that satisfy the condition $\left( A_{\rm in}(R) \in \, \left[\dfrac{n_{\rm in}}{\lambda_s}, \dfrac{n_{\rm in}+1}{\lambda_s}\right]\right)$ before solving \eqref{eq:P_N_in_numerical}. This expensive process should be repeated for different values of $n_{\rm in} \in [0,N]$ and over all time steps. This direct numerical solution is also prohibitive. 

Instead, to simplify the solution we note that the number of IoT devices inside  $\mathcal{B}_{\rm in}$ is a function of $R$, and combine \eqref{eq:P_d_given_B_in} and \eqref{eq:P_N_in_numerical} to get,
{\small 
\begin{align}
&P_{{\rm d}|{\rm int} }[k]= \nonumber \\
& \sum_{n=0}^{N} \int_{r\in \, \left\{ R: \, A_{\rm in}(R) \in \, \left[\dfrac{n}{\lambda_s}, \dfrac{n+1}{\lambda_s}\right] \right\} } \dfrac{2r}{\overline{R}_{u}^2[k]-\underline{R}_{u}^2[k]}  P_{ {\rm d}|n_{\rm in}(r) } dr, \\  \label{eq:P_d_given_B_in_2}
& =  \int_{\underline{R}_{u}[k]}^{\overline{R}_{u}[k]} \dfrac{2r}{\overline{R}_{u}^2[k]-\underline{R}_{u}^2[k]}  P_{ {\rm d}|n_{\rm in} (r)} dr,
\end{align} }
where $n_{\rm in}(r) = \left\lfloor \lambda_s A_{\rm int}(r) \right\rfloor$. Since $n_{\rm in}(r)$ cannot be solved in closed form, we approximate \eqref{eq:P_d_given_B_in_2} by 
\begin{align}\label{eq:P_d_given_B_in_approx}
P_{{\rm d}|{\rm int} }[k] \approx \sum_{i = 2}^{I} \dfrac{r_i^2 - r_{i-1}^2 }{\overline{R}_{u}^2[k]-\underline{R}_{u}^2[k]} P_{ {\rm d}|n_{\rm in} (r_i)},
\end{align}
where $r_i = \underline{R}_{u}[k] + \dfrac{\overline{R}_{u}[k]-\underline{R}_{u}[k] }{I}(i)$ and $I$ is an integer indicating the approximation accuracy. As $I \to \infty$, \eqref{eq:P_d_given_B_in_approx} converges to \eqref{eq:P_d_given_B_in_2}. 

In summary, we find the probability of fire detection at a given time step $k$ by solving \eqref{eq:P_uav} where $P_{{\rm d}|{\rm int} }[k]$ is expressed as in \eqref{eq:P_d_given_B_in_approx}. For each value of $r_i$, $n_{\rm in}(r_i)$ is computed by using \eqref{eq:A_in}-\eqref{eq:theta}. The value of $n_{\rm in}(r_i)$ is then used to obtain $P_{ {\rm d}|n_{\rm in} (r_i)}$ as expressed in \eqref{eq:Poisson_Binomial_2}. Similarly, the probability of false alarm at time step $k$ is obtained by solving \eqref{eq:Binomial}. The probabilities of fire detection and false alarm at time step $k$ are injected into the DTMC model. The procedure is repeated for the time steps $k=1, \cdots, K$. These steps are summarized in Algorithm~\ref{alg}. 

Before presenting the numerical analysis, we provide useful performance insights and discuss the UAV-IoT network design in the next section.

\begin{algorithm}[H]
	\caption{Wildfire detection probability}
	\begin{algorithmic}[1] \label{alg}	
		\STATE Initialize $I= \text{constant}$, $\boldsymbol{\pi}[0]=[1,0,0]$, $\pi_D[0]= \rho_D[0]=0$, and $ P_{NV}[0] = P_{NN}[0] = P_{VV}[0] = P_{NN}[0] = P_{ND}[0] = 0$. Define $\textbf{P}[k]$ as in \eqref{eq:transition_matrix}.
		\FOR{$k = 1:K$,}
		\FOR{$i = 1:I$,}
		\STATE $r_i= \underline{R}_{u}[k] + \dfrac{\overline{R}_{u}[k]-\underline{R}_{u}[k] }{I}(i)$,
		\STATE $ n_{\rm in}(r_i) = \lambda_s A_{\rm in}$ where $A_{\rm in}$ is as in  \eqref{eq:A_in},
		\STATE Calculate $P_{ {\rm d}|n_{\rm in} (r_i)}$ as in \eqref{eq:Poisson_Binomial_2},
		\ENDFOR
		\STATE $P_{{\rm d}|{\rm int} }[k] \approx \sum_{i = 2}^{I} \dfrac{r_i^2 - r_{i-1}^2 }{\overline{R}_{u}^2[k]-\underline{R}_{u}^2[k]} P_{ {\rm d}|n_{\rm in} (r_i)}$,
		\STATE $P_{\rm int}[k] = \dfrac{N_{u} A_u[k]}{A} $,
		\STATE $P_{\rm d}[k] = P_{\rm int}[k] P_{{\rm d}|{\rm int} }[k]$,
		\STATE $P_{\rm fa}[k] = \left(1-P_{\rm int}[k] \right) \left(1-\sum_{m=0}^{M-1} {{N} \choose m} \epsilon^{m}(1-\epsilon)^{{N}-m} \right)$,
		
		\STATE Update $\textbf{P}[k]$ based on $P_{\rm d}[k]$ and $P_{\rm fa}[k]$ as in \eqref{eq:P_NV}-\eqref{eq:P_VD}.
		\STATE $\boldsymbol{\pi}[k] = \boldsymbol{\pi}[k-1] \textbf{P}[k] $,
		\STATE $\pi_D[K] = \left[\boldsymbol{\pi}[k] \right]_3$,
		\STATE $\rho_D[k] = \pi_D[K] - \pi_D[K-1]$.
		\ENDFOR		
	\end{algorithmic}
\end{algorithm}

\section{Design and Performance Insights}
\label{sec:insigts}

In this section we discuss the UAV-IoT network design and provide insights on its performance for fire detection by studying a few special cases. In the previous section, derived detection probability so it is instructive to understand the roles of the system parameters on the wildfire detection probability.

First, as the number of UAVs $N_u$ increases, the performance strictly improves as the forest area covered by each UAV is reduced. This comes at an additional cost of more UAV deployments. Interestingly though, increasing the IoT device's density does not necessarily improve the detection probability. Note that $\lambda_s$ describes the trade off between network exploration and accurate diagnosis of explored area. While high $\lambda_s$ improves the detection/false alarm probabilities at any time step, it also implies that the UAV needs to spend more time at each hovering location and therefore less number of explored regions $K$ over a fixed mission time. Note that in practice, it is more cost effective to collect all IoT devices' measurements from a low density network than collecting a percentage of higher density network. However, the later approach is more reliable as it is tolerant to probable IoT device damage. 

Investment in the quality of the UAV and IoT devices can significantly improve the UAV-IoT wildfire detection system. A more agile UAV reduces the needed time to visit new hovering locations $T_{\rm travel}$ at a higher UAV cost $\omega_u$. It is also possible to increase fire detection distance $d_s$ at the IoT device, and decrease $\epsilon_s$ and $\epsilon_t$ by using more expensive IoT devices which can detect fire at higher distances, make less sensing errors and have more efficient energy harvesting equipment to supply higher transmission power, which effectively enlarges the UAV coverage region $R_{\rm hov}$.

For a fixed IoT device transmission power, a larger UAV coverage leads to high probability of non zero intersection between the sensor detection ring and the UAV coverage region at any time step ($P_{\rm int}[k]$). However, this comes at the cost of higher $\epsilon_t$ since higher channel path loss is expected. Further, the UAV hovering time is increased since more IoT devices are covered by the UAV at each time step which slows down the exploration of new regions. Finally, a high number of required positive flags $M$ decreases both the false alarm and detection probabilities. The error probability $\epsilon$ and the cost of false alarm, i.e., the verification time $T_{\rm vrf}$, play a major role in selecting optimal $M$ such that fire detection probability is maximized. 

Next, we study special UAV-IoT network designs to obtain insights on the system preference.
\begin{enumerate}
	\item $M=1$: By shifting to the verification mode once a positive flag is received at the UAV, the miss-detection probability is minimized and the false alarm probability is maximized. The expressions of these probabilities simplifies to, 
	\begin{align} \label{eq:fa_m=1}
	P_{\rm fa}[k] &= (1-P_{\rm int}[k])(1-(1-\epsilon)^N), \\ \label{eq:d_m=1}
	P_{ {\rm d}|n_{\rm in}  } &= 1 - \epsilon^{n_{\rm in}} (1 - \epsilon)^{n_{\rm out}},  
	\end{align}
	From \eqref{eq:fa_m=1}, the false alarm probability increases as the error probability and/or the number of covered IoT devices increase. While from \eqref{eq:d_m=1}, the detection probability increases as the number $N$ of covered IoT devices and/or the number $n_{\rm in}$ of IoT devices within $\mathcal{B}_{\rm in}$ increases. Note that as $\epsilon$ increases, $P_{ {\rm d}|n_{\rm in} }$ could increase or decrease based on the numbers $n_{\rm in}$ and $n_{\rm out}$. The case of $M=1$ significantly simplifies the computational complexity as described at the end of this section.
	Similar conclusions apply for $M>1$ with different rates of dependence. However, as $M$ increases, both the detection and false alarm probabilities decrease. Hence, $M$ should be carefully selected such that an optimal trade off is obtained.
	
	\item $\epsilon = 0$: In this case, $M=1$ is selected to maximize detection probability. Since any received positive flag at the UAV is correct, $P_{\rm fa}[k] =0$ and if $n_{\rm in}\geq 1$, $P_{ {\rm d}|n_{\rm in}  } =1$. Therefore, 
	\begin{align}
	P_{\rm d}[k] &= P_{\rm int}[k] \int_{\underline{R}_{u}[k] }^{\overline{R}_{u}[k]} \dfrac{2r}{\overline{R}_{u}^2[k]-\underline{R}_{u}^2[k]} P_{ {\rm d}|n_{\rm in} (r) } dr, \\
	&= P_{\rm int}[k] \left( \dfrac{\tilde{R}_{u}^2[k]-\underaccent{\tilde}{R}_{u}^2[k]}{\overline{R}_{u}^2[k]-\underline{R}_{u}^2[k]} \right),
	\end{align}
	where $\underaccent{\tilde}{R}_{u}[k]$ and $\tilde{R}_{u}[k]$ are the minimum and maximum radii of circles centered at origin, defining the region within which the UAV obtains $n_{\rm in} \geq 1$ on average. Eliminating the error probability enhances the UAV-IoT network performance significantly, for example, $P_{\rm d}[k] \approx P_{\rm int}[k]$ for dense IoT networks. To reduce $\epsilon$ more expensive IoT devices are required such that sensing is more accurate and more transmission power is available, and/or UAV coverage region is reduced to improve the transmission channel gain. 
	
%
%
\end{enumerate}

The computational complexity order for obtaining $\pi_D[K]$ and $\rho_D[K]$ is $\mathcal{O}(IKM^2)$ where $I$ is the number of circle radii used to approximate the number of IoT devices in $\mathcal{B}_{\rm in}$. Equations \eqref{eq:detection_at_k_1}-\eqref{eq:P_fa}, \eqref{eq:P_uav}, \eqref{eq:P_d_given_B_in}, and \eqref{eq:P_d_given_B_in_approx} are solved at each time step, $k \, \forall k \in \{1, \cdots, K\}$. In each time iteration, the most complex operation which is solving the Poisson Binomial distribution in (26) is solved $I$ times. The computational cost to solve the Poisson Binomial distribution is $\mathcal{O}(M^2)$. Note that to obtain $\pi_D[K]$, all the values $\pi_D[k] \, \forall k \in \{1, \cdots, K\}$ are obtained. The values of $\rho_D[k] \, \forall k \in \{1, \cdots, K\} $ are simply obtained as $ \rho_D[k] = \pi_D[k] - \pi_D[k-1]$. In the case $M=1$, the computational complexity simplifies to $\mathcal{O}(IK)$ since the Binomial distribution and the Poisson Binomial distribution computations are reduced as given in \eqref{eq:fa_m=1} and \eqref{eq:d_m=1}. 

The optimization problems in Section~\ref{sec:problem_stat} are NP-hard, therefore, it is not possible to solve them analytically in polynomial time. Since we are optimizing over only three variables (i.e. $N_u$, $\lambda_s$ and $M$), a simple search algorithm can be utilized to solve the fire detection and fire fighting problems. Note that for a given system budget $\zeta$ and IoT devices cost, the number of UAVs is maximized as, 
\begin{align}
N_u =\dfrac{\zeta -\omega_{s} N_{s}}{\omega_u },
\end{align}
to maximize fire detection probability. To solve P1, we search over $\lambda_s$ and $M$ for the maximum $\pi_D[K]$. To solve P2, we search over different system budgets for the minimum wildfire losses.

\section{Numerical Results}
\label{sec:numerical_results}

\begin{figure}
	\centering
	\begin{tikzpicture}[thick,scale=0.7, every node/.style={scale=0.7}]
	\draw (0, 0) node[inner sep=0] {	\includegraphics[trim={0cm 0cm  0cm 0cm},clip, width=1.4 \linewidth]{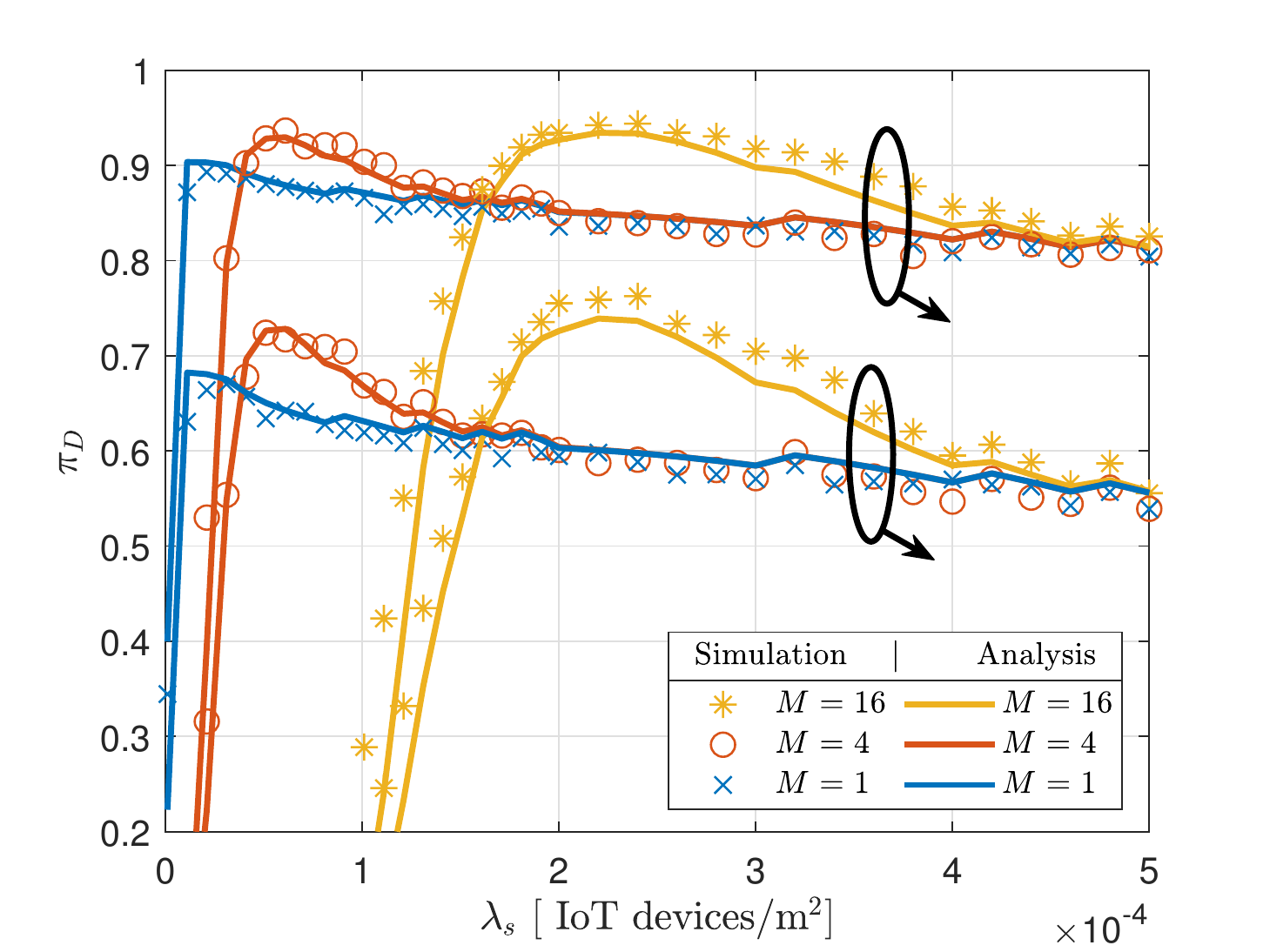}};
	\draw (3.7,1.4) node {{$N_u = 20$}};		
	\draw (3.3,-1.1) node {{$N_u = 10$}};	
	\end{tikzpicture}
	\caption{Detection probability versus device density $\lambda_s$ for different values of $M$ and $N_u$.} \label{fig:lambda_s}
\end{figure}

\begin{figure}
	\centering
	\begin{tikzpicture}[thick,scale=0.7, every node/.style={scale=0.7}]
	\draw (0, 0) node[inner sep=0] {	\includegraphics[trim={0cm 0cm  0cm 0cm},clip, width=1.4 \linewidth]{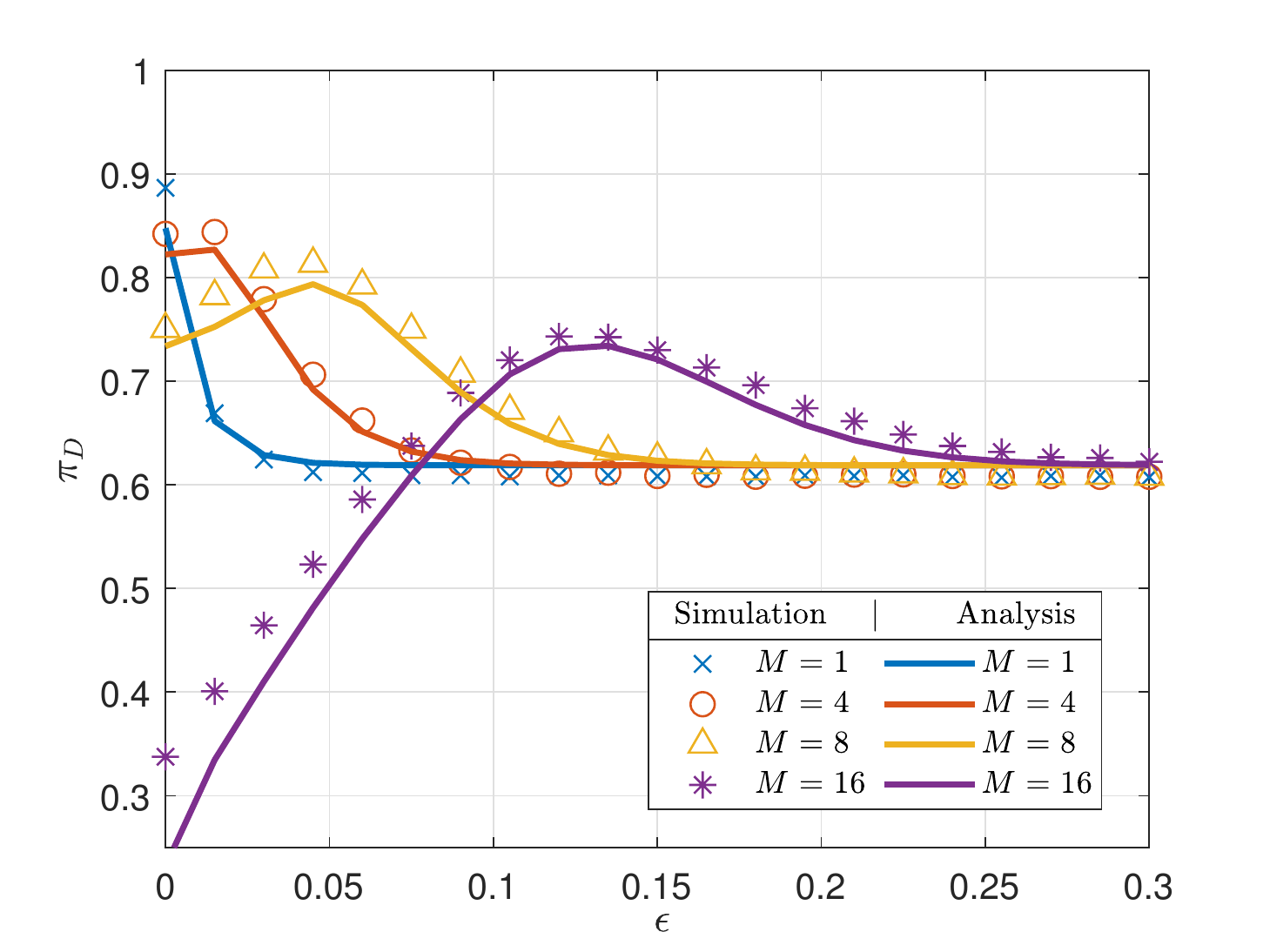}};
	\end{tikzpicture}
	\caption{Detection probability against $\epsilon$.} \label{fig:p_e}
\end{figure}

\begin{figure}
	\centering
	\begin{tikzpicture}[thick,scale=0.7, every node/.style={scale=0.7}]
	\draw (0, 0) node[inner sep=0] {	\includegraphics[trim={0cm 0cm  0cm 0cm},clip, width=1.33 \linewidth]{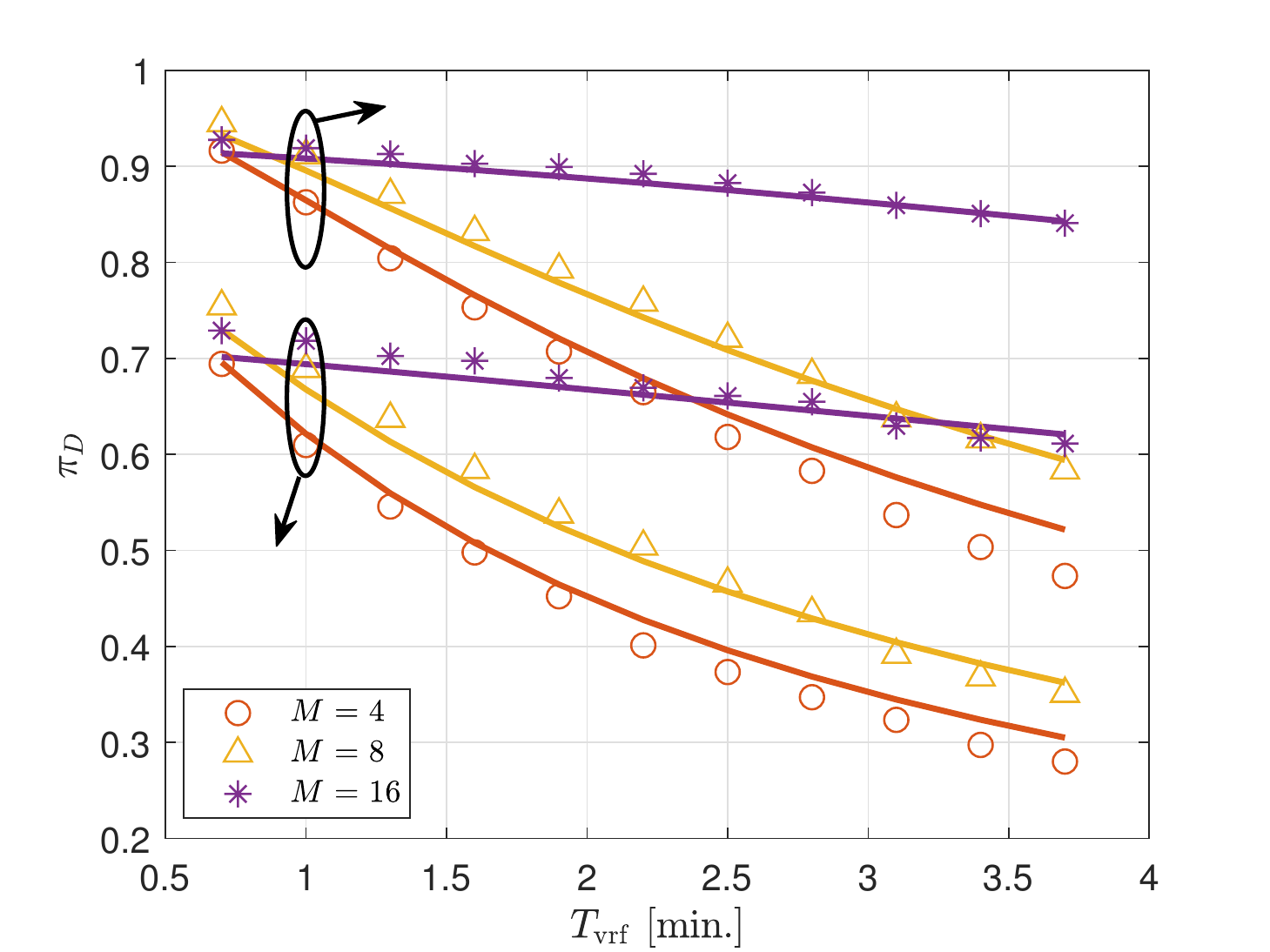}};
	\draw (-2.2, -2.4) node[inner sep=0] {\includegraphics[trim={0cm 0cm  0cm 0cm},clip, width=.48 \linewidth]{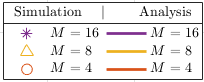}};
	\draw (-1.2,3.3) node {{$N_u = 20$}};		
	\draw (-3.3,-.9) node {{$N_u = 10$}};	
	\end{tikzpicture}
	\caption{The effect of the fire verification time on the detection probability.} \label{fig:T_vrf}
\end{figure}

\begin{figure}
	\centering
	\begin{tikzpicture}[thick,scale=0.7, every node/.style={scale=0.7}]
	\draw (0, 0) node[inner sep=0] {	\includegraphics[trim={0cm 0cm  0cm 0cm},clip, width=1.33 \linewidth]{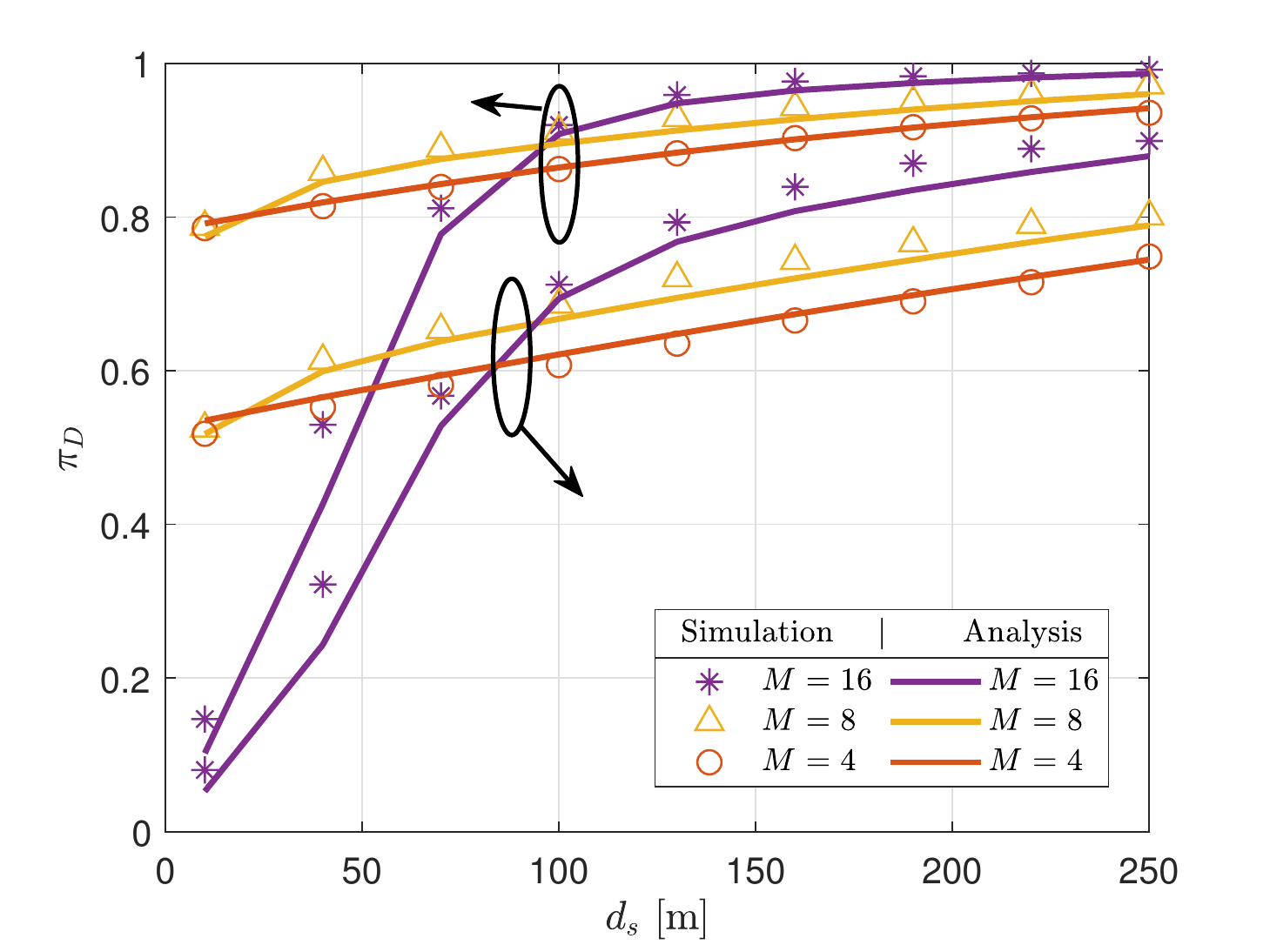}};
	\draw (-2.5,3.3) node {{$N_u = 20$}};		
	\draw (-0.3,-.5) node {{$N_u = 10$}};	
	\end{tikzpicture}
	\caption{Detection probability against IoT device's detection range.} \label{fig:d_s}
\end{figure}

\begin{figure}
	\centering
	\begin{tikzpicture}[thick,scale=0.67, every node/.style={scale=0.67}]
	\draw (0, 0) node[inner sep=0] {	\includegraphics[trim={0cm 0cm  0cm 0cm},clip, width=1.33 \linewidth]{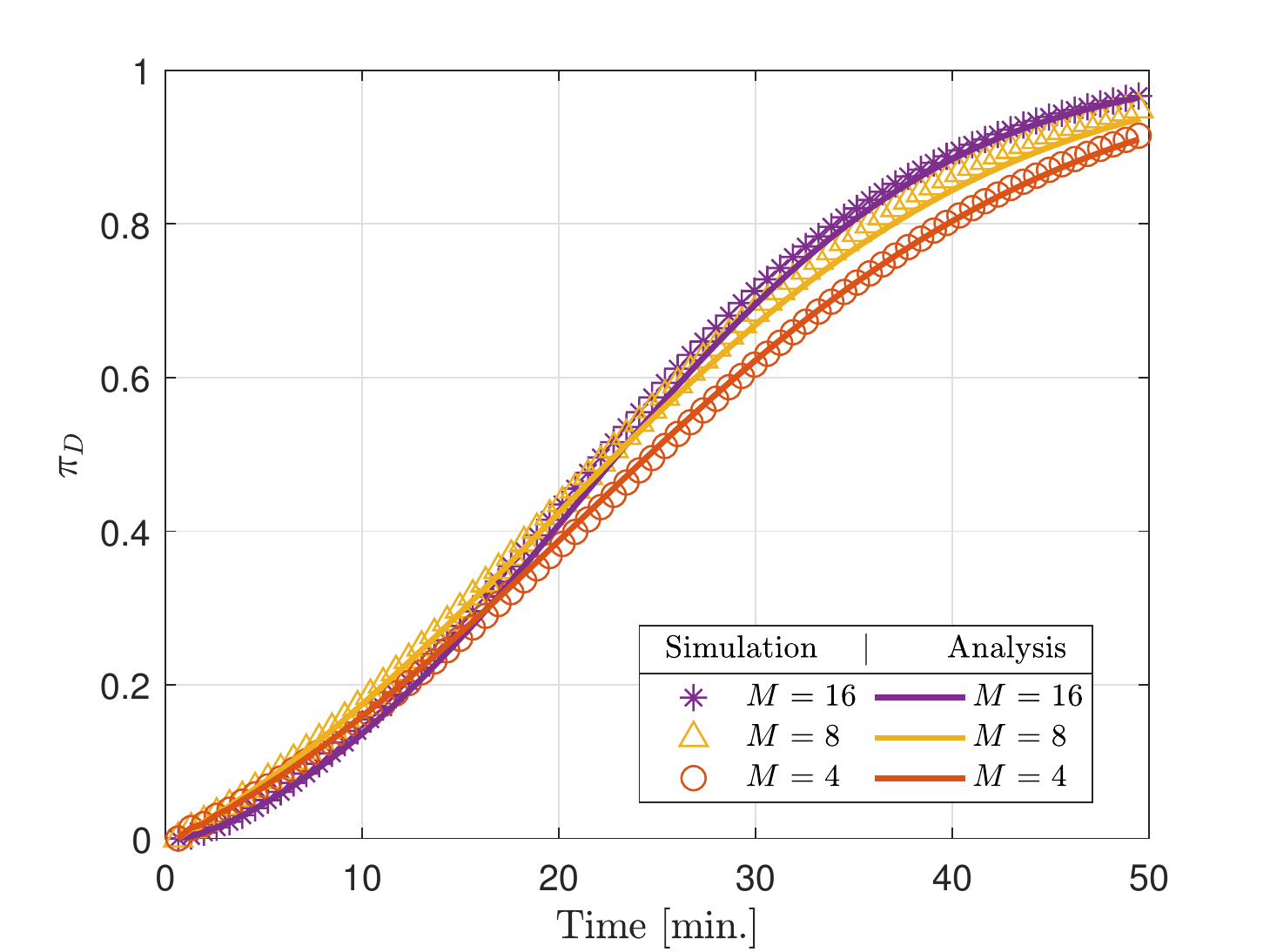}};
	\end{tikzpicture}
	\caption{Detection probability by different times starting from the fire ignition.} \label{fig:pi}
\end{figure}

\begin{figure}
	\centering
	\begin{tikzpicture}[thick,scale=0.67, every node/.style={scale=0.67}]
	\draw (0, 0) node[inner sep=0] {	\includegraphics[trim={0cm 0cm  0cm 0cm},clip, width=1.33 \linewidth]{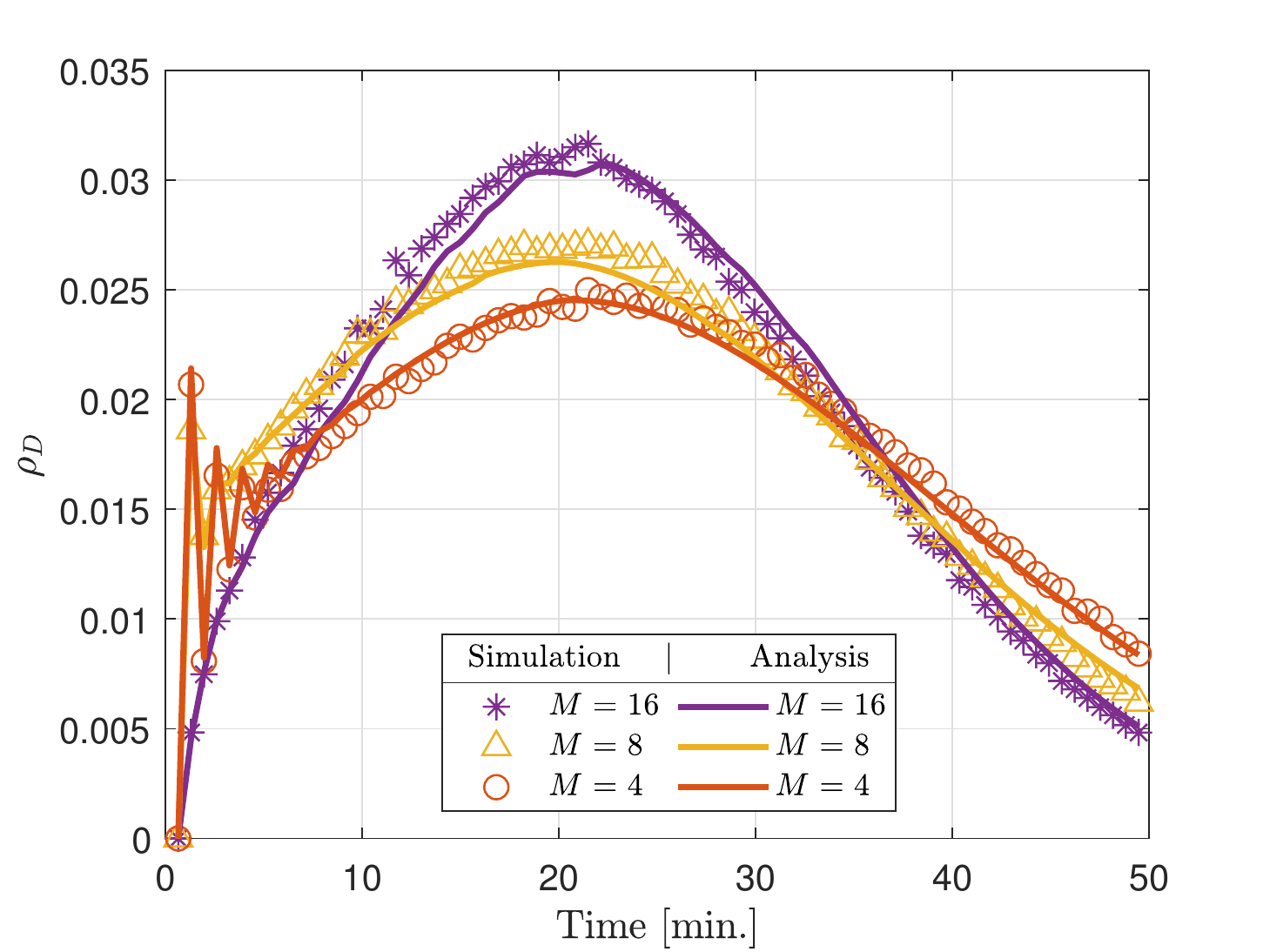}};
	\end{tikzpicture}
	\caption{Detection probability at different times starting from the fire ignition.} \label{fig:rho}
\end{figure}

In this section, we validate our analysis, provide insightful performance figures, and show numerical solutions for the wildfire detection probability maximization and the wildfire losses minimization problems. To validate the model, the mathematical analysis is compared with independent Monte Carlo simulation. At each iteration in the simulation environment, IoT devices are deployed randomly over the forest area and a fire starts at a random location at time step $k=0$. At each time step, the UAVs visit a new location and collect measurements from covered sensors to detect the fire until the fire is detected or the critical time, $T_f$ is reached. The fire detection probability is calculated by dividing the number of iterations where the fire was detected by the UAVs over the total number of iterations. In Fig.~\ref{fig:lambda_s}, \ref{fig:p_e}-\ref{fig:rho}, the lines represent the analysis results while the markers represent the simulation results.

Let's consider $N$ IoT devices distributed uniformly over a forest of area, $A = 20 $ km $\times 20 $ km $ = 400$ km$^2$ (which is equivalent to 56,022 football fields) with density $\lambda_s = 180$ IoT devices per km$^2$. Assume that a fire is ignited at a random location such that it spreads in all directions at a constant rate of spread (ROS) of $v=20$ m/min. A number of UAVs, $N_u= 10$, are covering the forest such that each UAV is responsible for detecting the fire within an equal portion of the forest. The detection probability needs to be maximized so that a fire is detected before the critical time $T_f= 30$ mins. Unless otherwise mentioned, the system parameters in Table~\ref{table:par} are assumed throughout this section. 

{\small 
\renewcommand{\arraystretch}{0.85}
\begin{table}
	\caption{Default system parameters.}	\label{table:par} 
	\centering
	\begin{tabular}
		{| p{.4cm} | p{1.6cm}  ||  p{.5cm} | p{1.2cm} ||  p{.8cm} | p{1.5cm} |} 
		\hline
		Par. & Value & Par. & Value & Par. & Value \\ [0.5ex] 
		\hline\hline
		$\lambda_s$ & $180$ IoT devices/km$^2$ &$N_u$ & $10$ & $M$& $\{1,4,8,16\}$ \\ 
		\hline
		$A$ & $20\times20$ km$^2$ &$v$ & $20$ [m/min] & $d_s$& $100$ [m] \\ 
		\hline
		$\epsilon$ & $0.1$ & $R_{\rm hov}$ & $400$ [m]  & $T_{\rm travel}$ & $0.5$ [min] \\
		\hline
		$ T_{\rm obs}$ & $0.1$ [s] & $\beta$ & $1$ & $\zeta$  & $10\times10^6$   \\
		\hline
		$T_{\rm vrf}$ & $1$ [min] & $T_f$ & $ 30 $ [min] & $T_D$ & $30$ [min]  \\
		\hline
		$\omega_s$ & $1$ & $\omega_u$ & $ 1000$ & $\omega_D(t)$ & $10000t^2$   \\
		\hline
	\end{tabular}
\end{table}
}

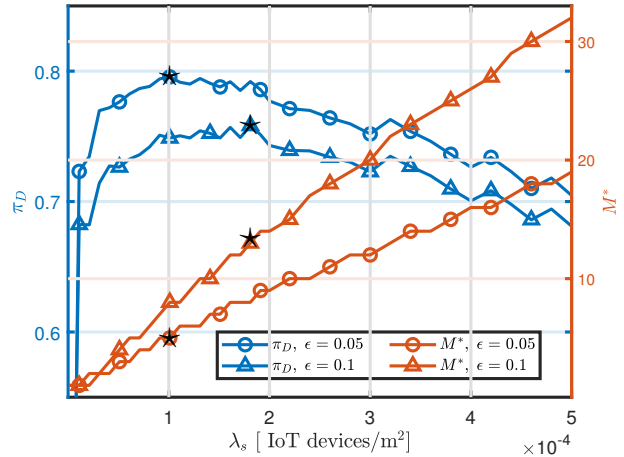
\begin{figure}
	\centering
	\begin{tikzpicture}[thick,scale=0.65, every node/.style={scale=0.65}]
	\draw (0, 0) node[inner sep=0] {	\includegraphics[width=1.5 \linewidth]{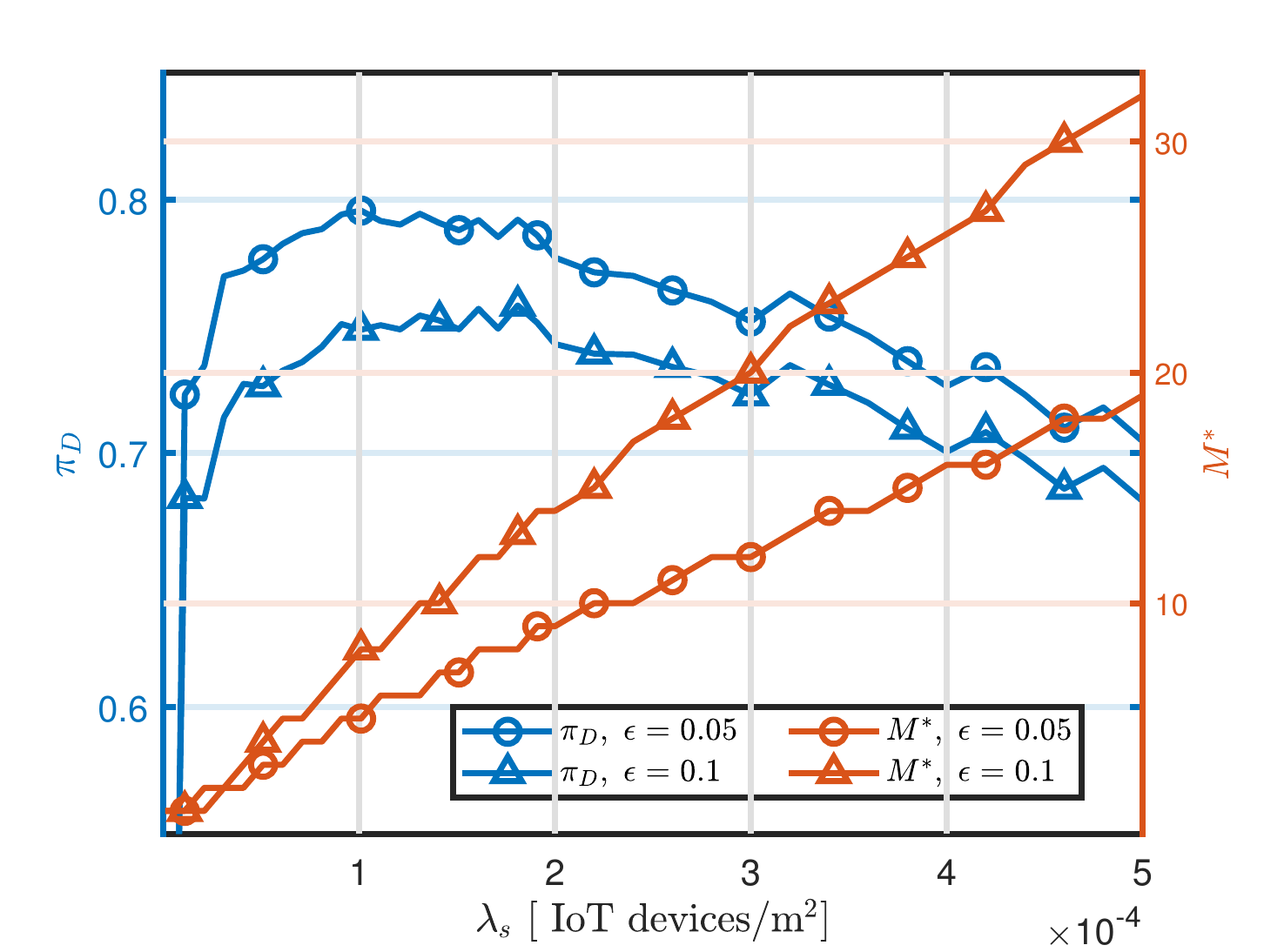}};
	\draw (-2.85,2.8) node {\Huge{$\star$}};	
	\draw (-1.2,1.8) node {\Huge{$\star$}};		
	\draw (-1.2,-.5) node {\Huge{$\star$}};		
	\draw (-2.85,-2.55) node {\Huge{$\star$}};	
	\end{tikzpicture}
	\caption{Detection probability against IoT device density $\lambda_s$ with optimal number of positive flags $M^*$ and $\epsilon = \{0.05, 0.1\}$. The selected $M^*$ against $\lambda_s$ is shown in the right $y$-axis. Stars ($\star$) denote the optimal pair of $\lambda_s$ and $M$ such that $\pi_D$ is maximized for given $\epsilon$.} \label{fig:new}
\end{figure}


We first show the detection performance for several values of $\lambda_s$, $M$ and $N_u$ in Fig.~\ref{fig:lambda_s}. As the figure shows, the probability of fire detection increases with the IoT devices density until $\lambda_s$ reaches an optimal value after which the fire detection probability decreases as  $\lambda_s$ increases. The reason behind the decreased performance of higher IoT devices densities is that the UAV spends a long period to collect a fixed percentage of the covered IoT devices which adds to the hovering time $T$. As a result, the UAV visits less number of locations over the critical time, $T_f =30$ min. This is good news in practice, since it implies that the best performance may be achieved for a moderate sensor density. Note that in practice $\beta$ can be adjusted such that less percentage of covered IoT data is collected at the UAV. This approach adds to the system cost as more number of IoT devices are deployed but enhances the system reliability against damaged or uncharged IoT devices. The coverage probability monotonically increases with the number of UAVs. Also we notice that as $M$ increases, the optimal $\lambda_s$ increases to obtain enough observations at the UAV and avoid miss-detection. Note that the gap between the simulation results and the analysis is due to the floor function approximation, $n_{\rm in}(r) = \left\lfloor \lambda_s A_{\rm in}(r) \right\rfloor$, the limitation in the number of summation terms in \eqref{eq:P_d_given_B_in_approx}, and the assumption that exactly $N$ IoT devices are covered by the UAV at any time while the number in the simulation is Poisson distributed with average $N$ IoT devices. In Fig. \ref{fig:new}, the maximum detection probability is shown against the IoT device density by selecting the optimal number of required positive flags denoted as $M^*$ for sensing and transmission error $\epsilon = \{0.05, 0.1\}$. The optimal number of required positive flags $M^*$ is shown in right $y$-axis. Note how $M^*$ increases as $\lambda_s$ increase in an approximately linear fashion with a slope dependent on the error probability $\epsilon$. Also, observe that there is an optimal IoT device density, denoted by Star ($\star$) in the figure, beyond which the detection probability starts to decrease.

In Fig.~\ref{fig:p_e}, the detection probability $\pi_D$ is evaluated versus $\epsilon$ for different values of $M$. As $\epsilon$ is low the false alarm probability is low and therefore $M$ is minimized to guarantee highest detection probability. As $\epsilon$ increases, higher $M$ will give a higher $\pi_D$ such that a better miss-detection / false alarm trade off is maintained. When $M$ is high and $\epsilon$ is low, the detection probability is very low. This is because the UAV covers $\lfloor 400^2\pi\times180\times10^{-6} \rfloor= 90$ devices on average and needs to collect a large number of positive flags. If $M=16$, the UAV should be placed such that at least $16$ IoT devices fall inside $\mathcal{B}_{\rm in}$ when $\epsilon = 0$. Given $d_s = 100$ this is very less likely, leading to a high miss-detection rate. When the UAV coverage circle is centered at the middle of the sensing detection ring, the average number of covered detecting devices converges to $n_{\rm in} \approx \lambda_s 2d_s R_{\rm hov} = 14.4$, which makes it extremely difficult to collect $16$ positive flags from sensors. All values converge to $\pi_D = 0.6 $ as $\epsilon$  becomes high. This is because verification mode is reached in each hovering period by false alarm with high probability. This detection probability can be slightly improved by choosing $M>16$ or significantly improve at a higher system budget by increasing the number of UAVs. 

While verification allows the UAV to examine the precision of this detection, the verification time also acts as a penalty in case of a false alarm, such that $T_{\rm vrf}$ is wasted every time false alarm is declared at the UAV. Hence, higher $M$ values are preferred for high $T_{\rm vrf}$, since this reduce the false alarm probability, as Fig.~\ref{fig:T_vrf} shows.  

As the IoT device's detection range increases the wildfire detection probability increases. This decreases the miss-detection probability, allowing us to increase $M$ to decrease the false alarm probability, while maintaining good detection performance. Thus, it is more desirable to choose a higher number $M$ of positive flags to declare fire at the UAV when $d_s$ is high, as Fig.~\ref{fig:d_s} shows.

In Fig.~\ref{fig:pi} and \ref{fig:rho}, the wildfire detection probabilities $\pi_D$ and $\rho_D$ are shown at the end of each hovering period $k, \; k \in \{1, \cdots, K\}$ where $K = \left\lfloor \dfrac{T_f}{T } \right\rfloor$ with $T_f=30$ min. and 
\begin{align}
T = N T_{\rm obs} + T_{\rm travel} = (\lambda_s \pi R_{\rm hov}^2 ) T_{\rm obs} + T_{\rm travel} = 0.65 \text{ min.}
\end{align}
The probability $\pi_D$ increases with time until eventually it approaches one. Fig.~\ref{fig:pi} indicates the probability of detecting the wildfire by the time indicated by the $x$-axis. On the other hand, Fig.~\ref{fig:rho} shows the probability of wildfire detection exactly at the time indicated in the $x$-axis. The detection at time step $k$ increases with time since the fire size increases with time, until a point where the detection at time $k$ starts to decrease because there is low chance the fire survives until this time without being detected. Again, different values of $M$ offer trade offs between wildfire miss-detection and false alarm probabilities. Fig.~\ref{fig:rho} is essential to obtain the expected wildfire losses cost.

\begin{figure}
	\centering
	\begin{tikzpicture}[thick,scale=0.66, every node/.style={scale=0.66}]
	\draw (0, 0) node[inner sep=0] {	\includegraphics[trim={0cm 0cm  0cm 0cm},clip, width=1.33 \linewidth]{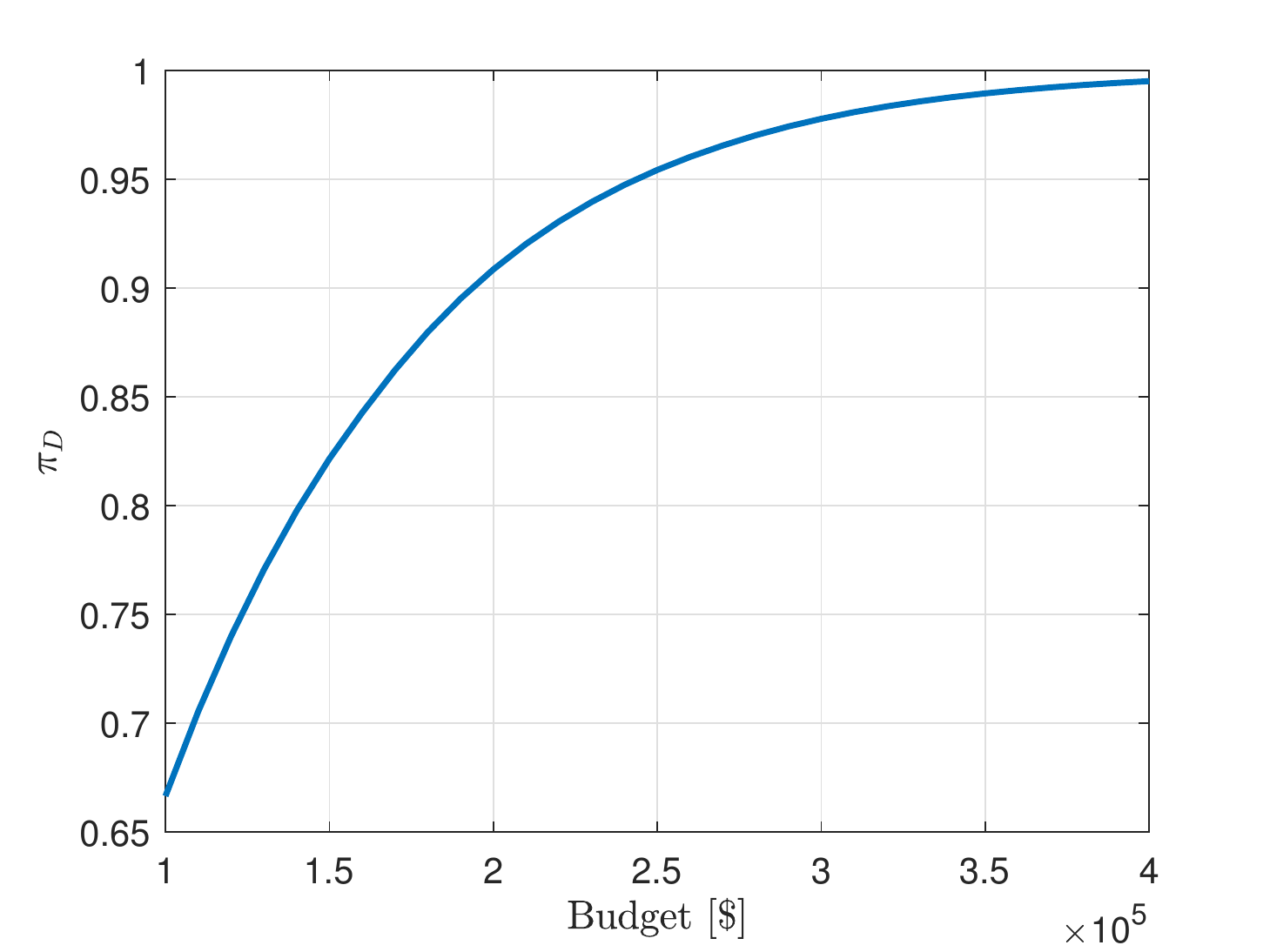}};
	\end{tikzpicture}
	\caption{Maximum wildfire detection probability for optimal $\lambda_s$, $M$ and $N_u$ values against system budget.} \label{fig:op1A}
\end{figure}

\begin{figure}
	\centering
	\begin{subfigure}{.42\linewidth}
		\centering
		\begin{tikzpicture}[thick,scale=1, every node/.style={scale=1}]
		\draw (0, 0) node[inner sep=0] {	\includegraphics[clip, width=1.1 \linewidth]{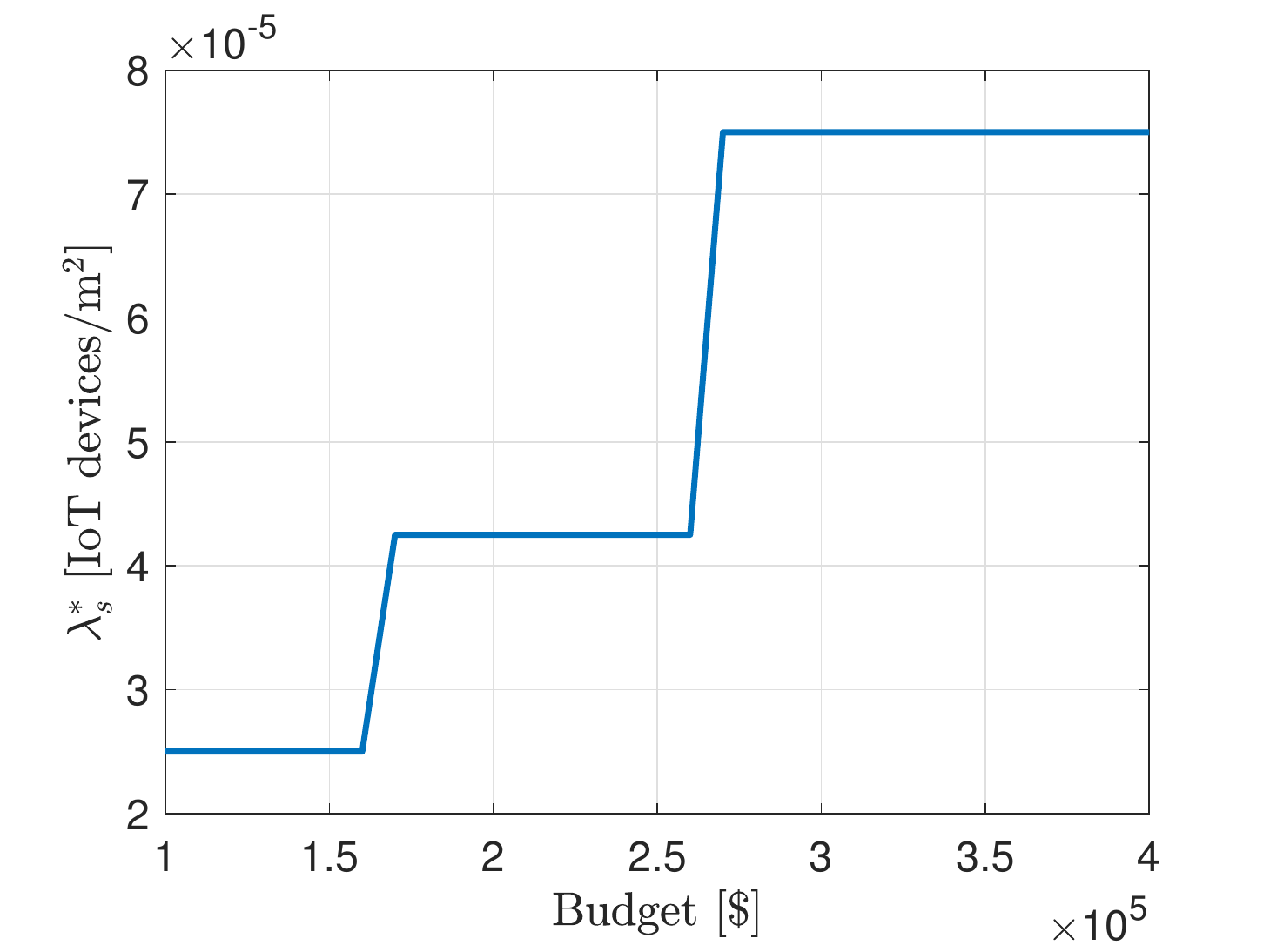}};	
		\end{tikzpicture}
	\end{subfigure}
	\begin{subfigure}{.42\linewidth}
	\centering
	\begin{tikzpicture}[thick,scale=1, every node/.style={scale=1}]
	\draw (0, 0) node[inner sep=0] {	\includegraphics[clip, width=1.1 \linewidth]{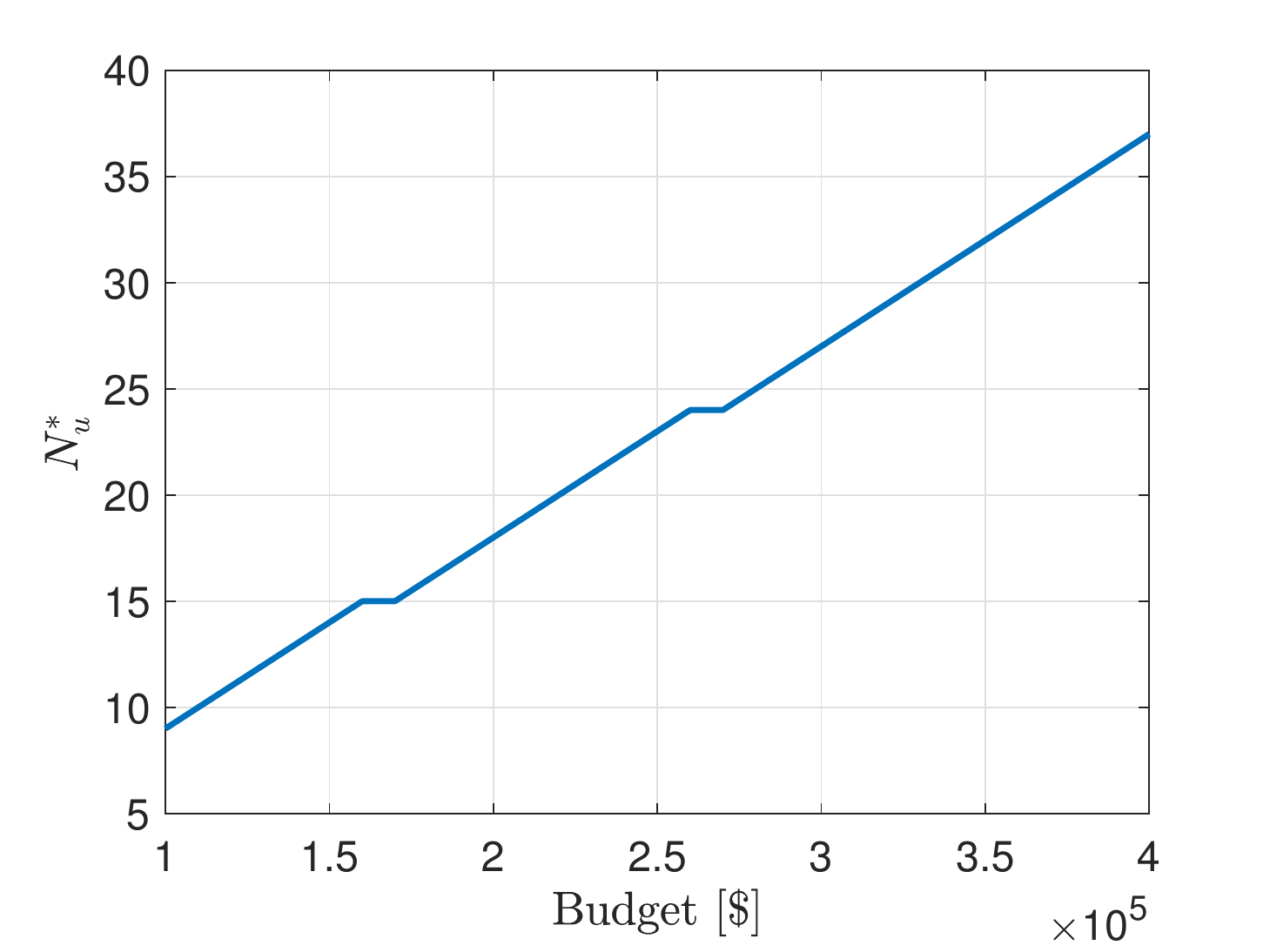}};	
	\end{tikzpicture}
	\end{subfigure}
	\begin{subfigure}{.42\linewidth}
		\centering
		\begin{tikzpicture}[thick,scale=1, every node/.style={scale=1}]
		\draw (0, 0) node[inner sep=0] {	\includegraphics[clip, width=1.1 \linewidth]{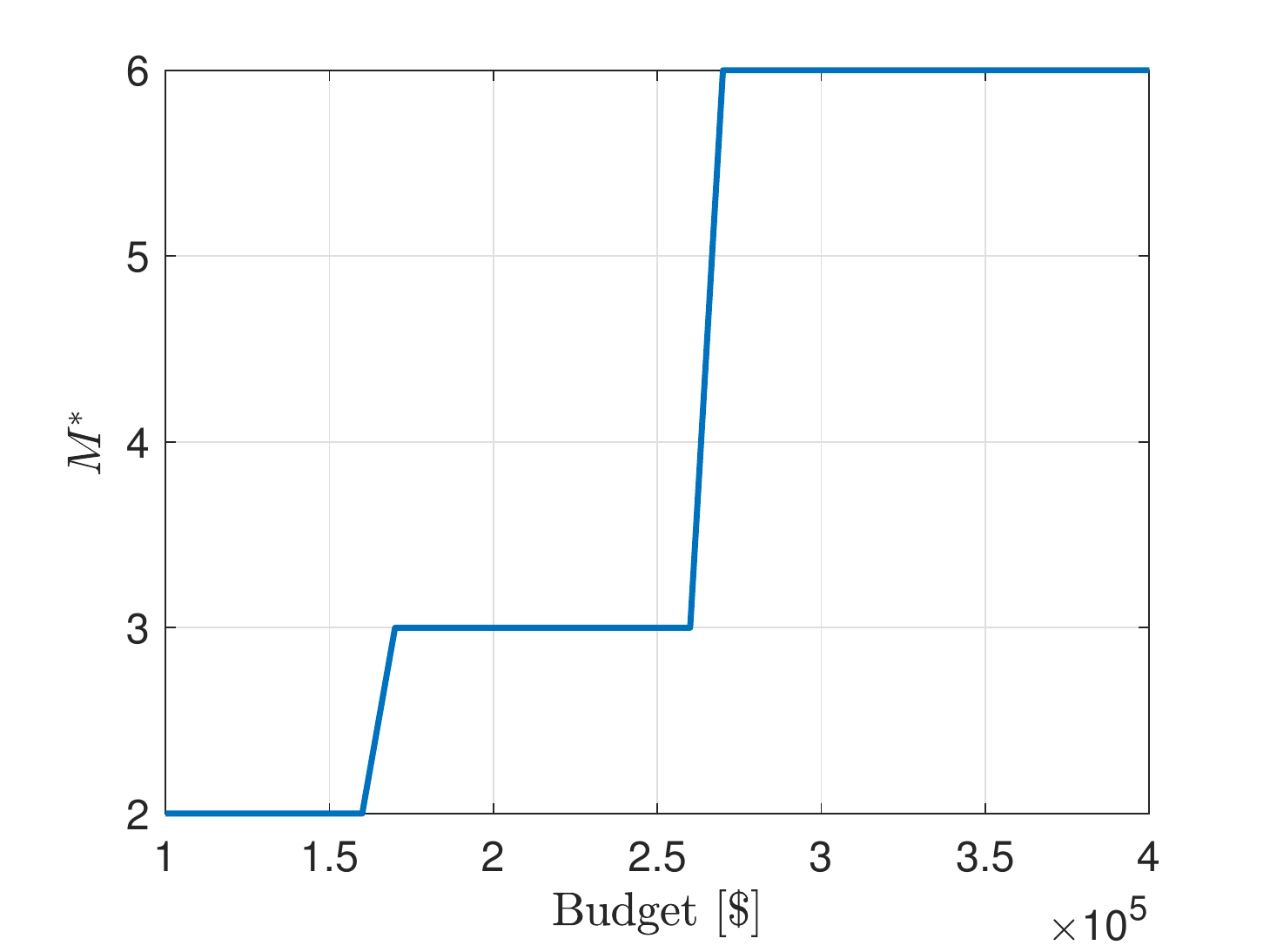}};	
		\end{tikzpicture}
	\end{subfigure}
	\caption{Optimal $\lambda_s$, $N_u$ and $M$ values for maximum wildfire detection probability against system budget.}  \label{fig:op1B}
\end{figure}

\begin{figure}
	\centering
	\begin{tikzpicture}[thick,scale=0.66, every node/.style={scale=0.66}]
	\draw (0, 0) node[inner sep=0] {	\includegraphics[trim={0cm 0cm  0cm 0cm},clip, width=1.33 \linewidth]{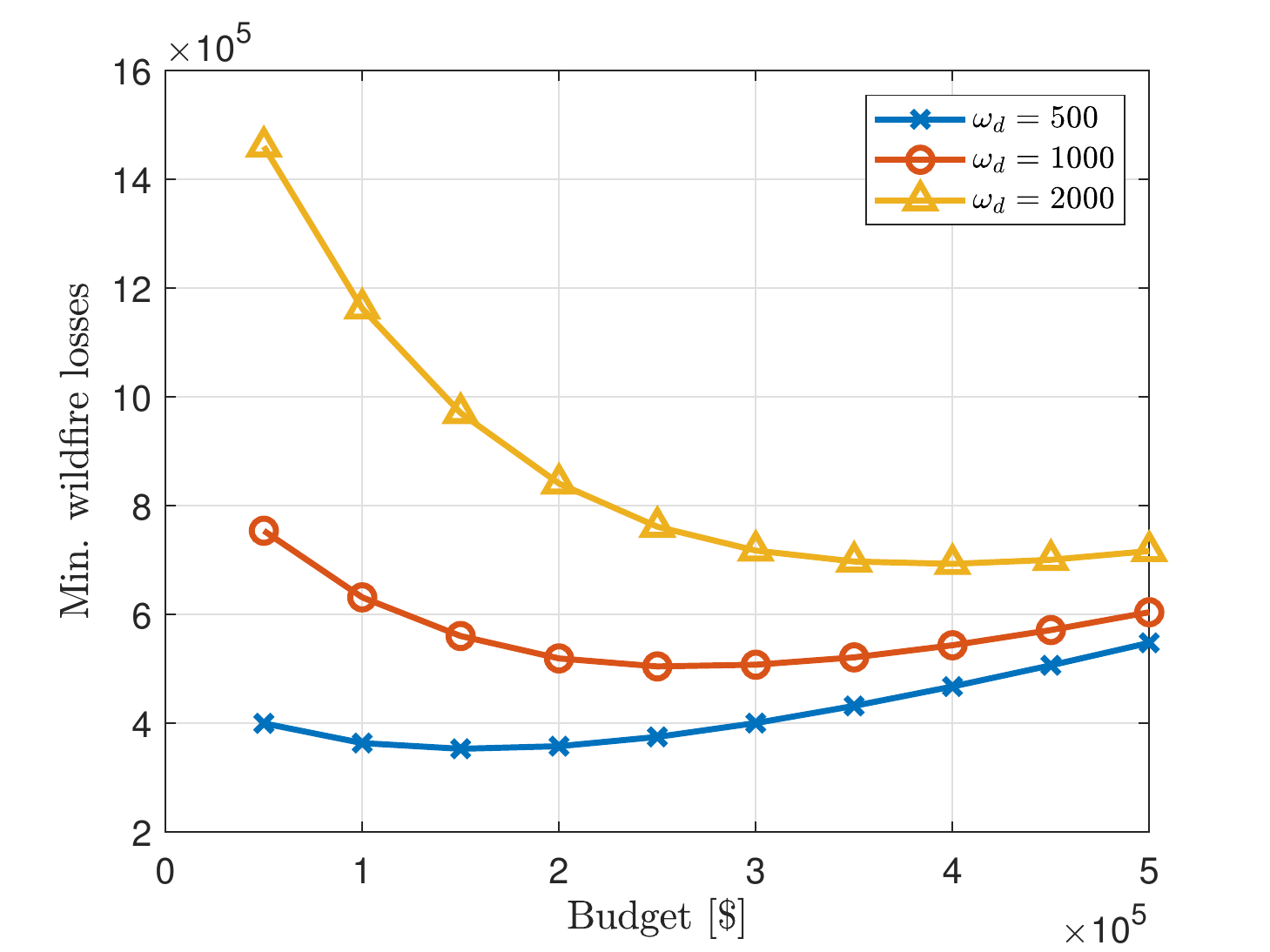}};
	\draw (-1.6, -2.6) node {\Huge \textbf{*} };	
	\draw (.2, -1.83) node {\Huge \textbf{*} };
	\draw (2.9, -0.9) node {\Huge \textbf{*} };
	\end{tikzpicture}
	\caption{Minimum wildfire losses for optimal $\lambda_s$, $M$ and $N_u$ values against system budget.} \label{fig:op2A}
\end{figure}
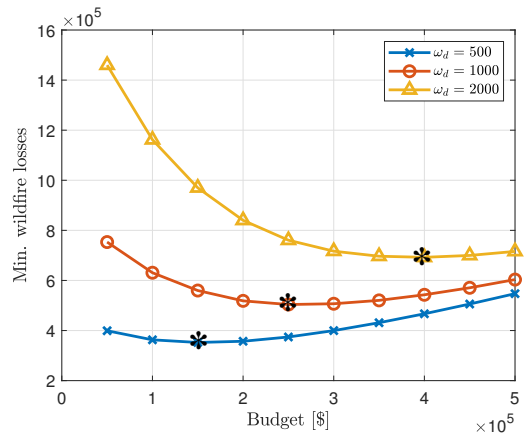

\begin{figure}
	\centering
	\begin{subfigure}{.42\linewidth}
		\centering
		\begin{tikzpicture}[thick,scale=1, every node/.style={scale=1}]
		\draw (0, 0) node[inner sep=0] {	\includegraphics[clip, width=1.1 \linewidth]{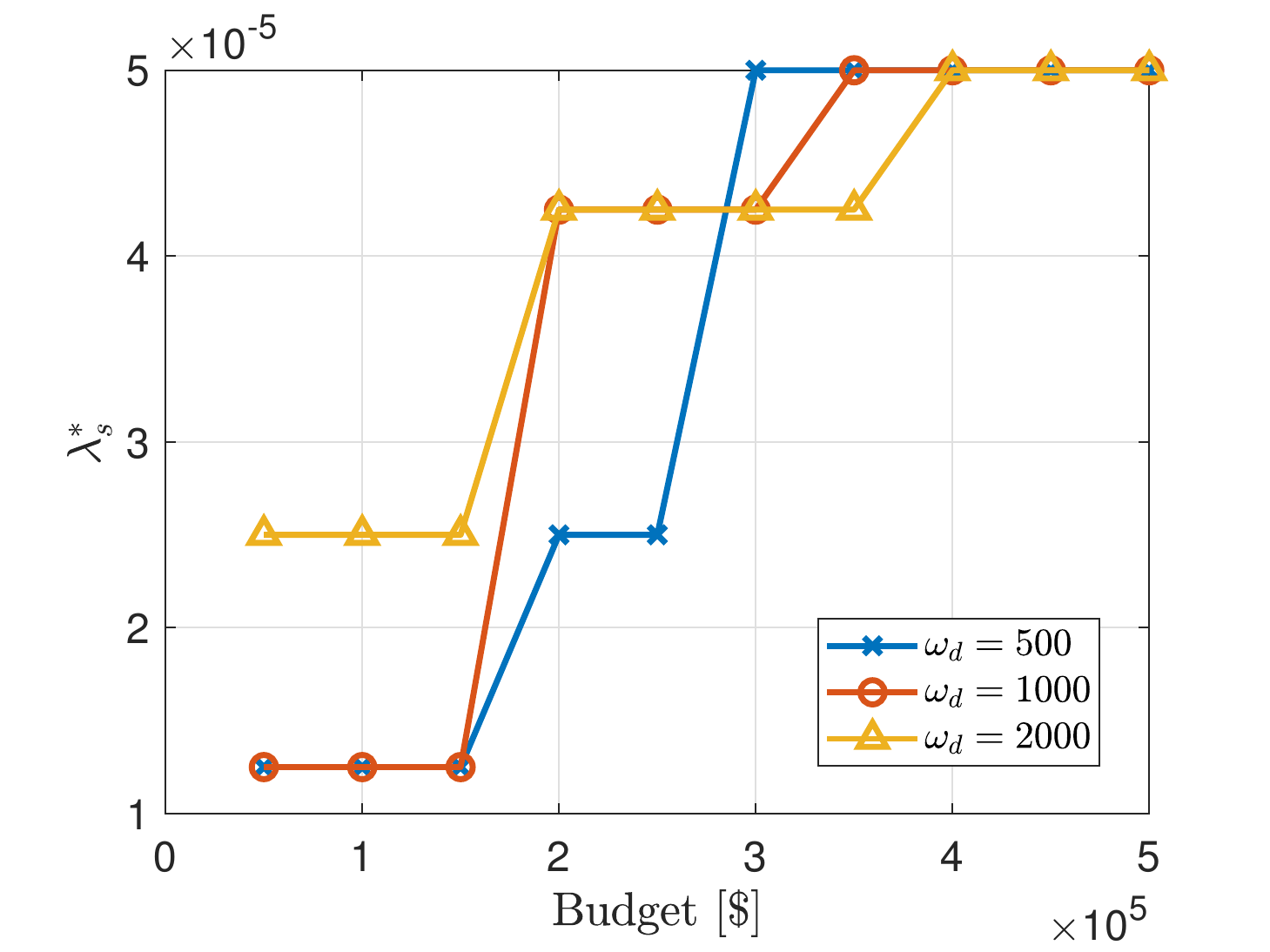}};	
		\end{tikzpicture}
	\end{subfigure}
	\begin{subfigure}{.42\linewidth}
		\centering
		\begin{tikzpicture}[thick,scale=1, every node/.style={scale=1}]
		\draw (0, 0) node[inner sep=0] {	\includegraphics[clip, width=1.1 \linewidth]{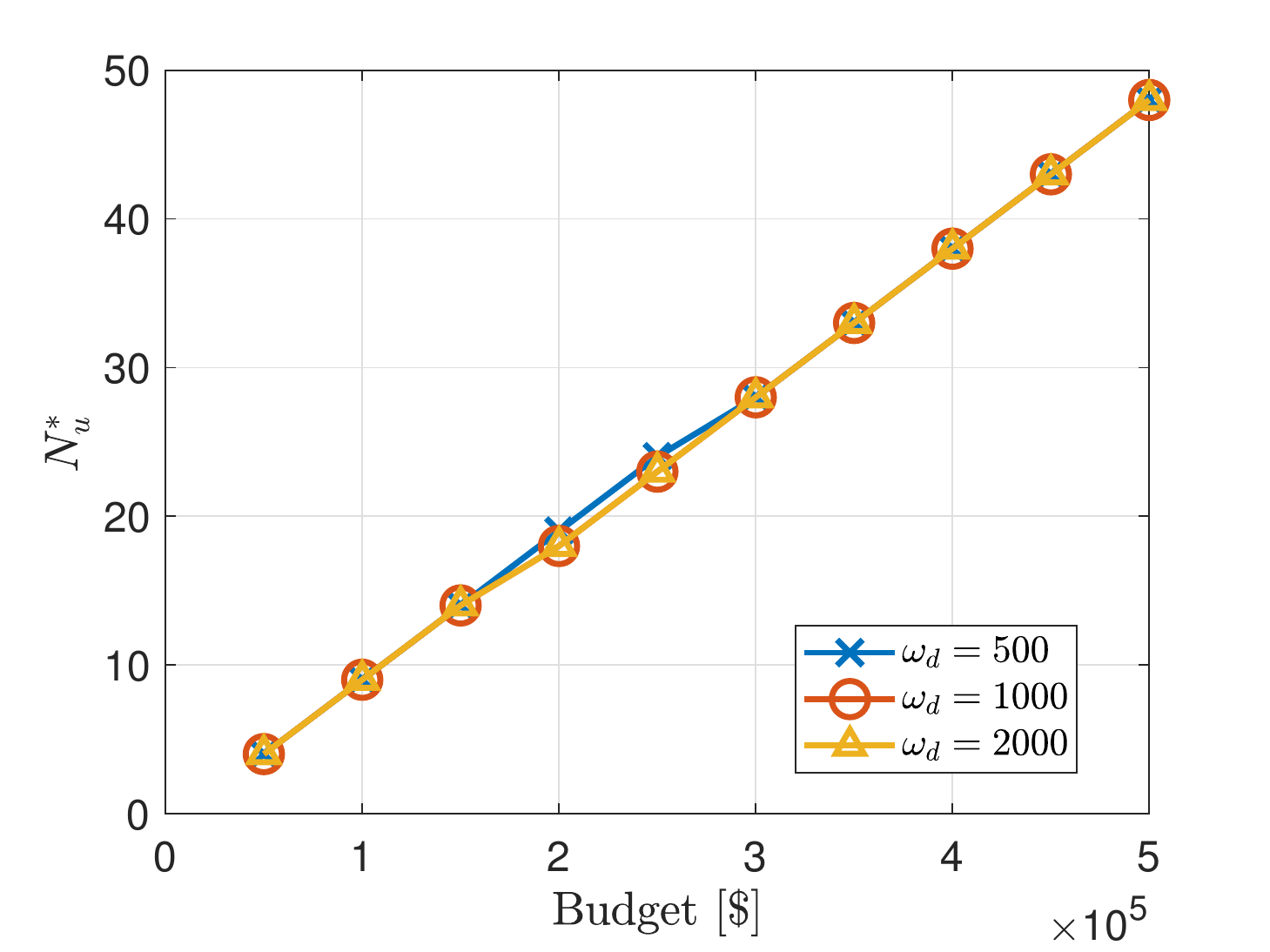}};	
		\end{tikzpicture}
	\end{subfigure}
	\begin{subfigure}{.42\linewidth}
		\centering
		\begin{tikzpicture}[thick,scale=1, every node/.style={scale=1}]
		\draw (0, 0) node[inner sep=0] {	\includegraphics[clip, width=1.1 \linewidth]{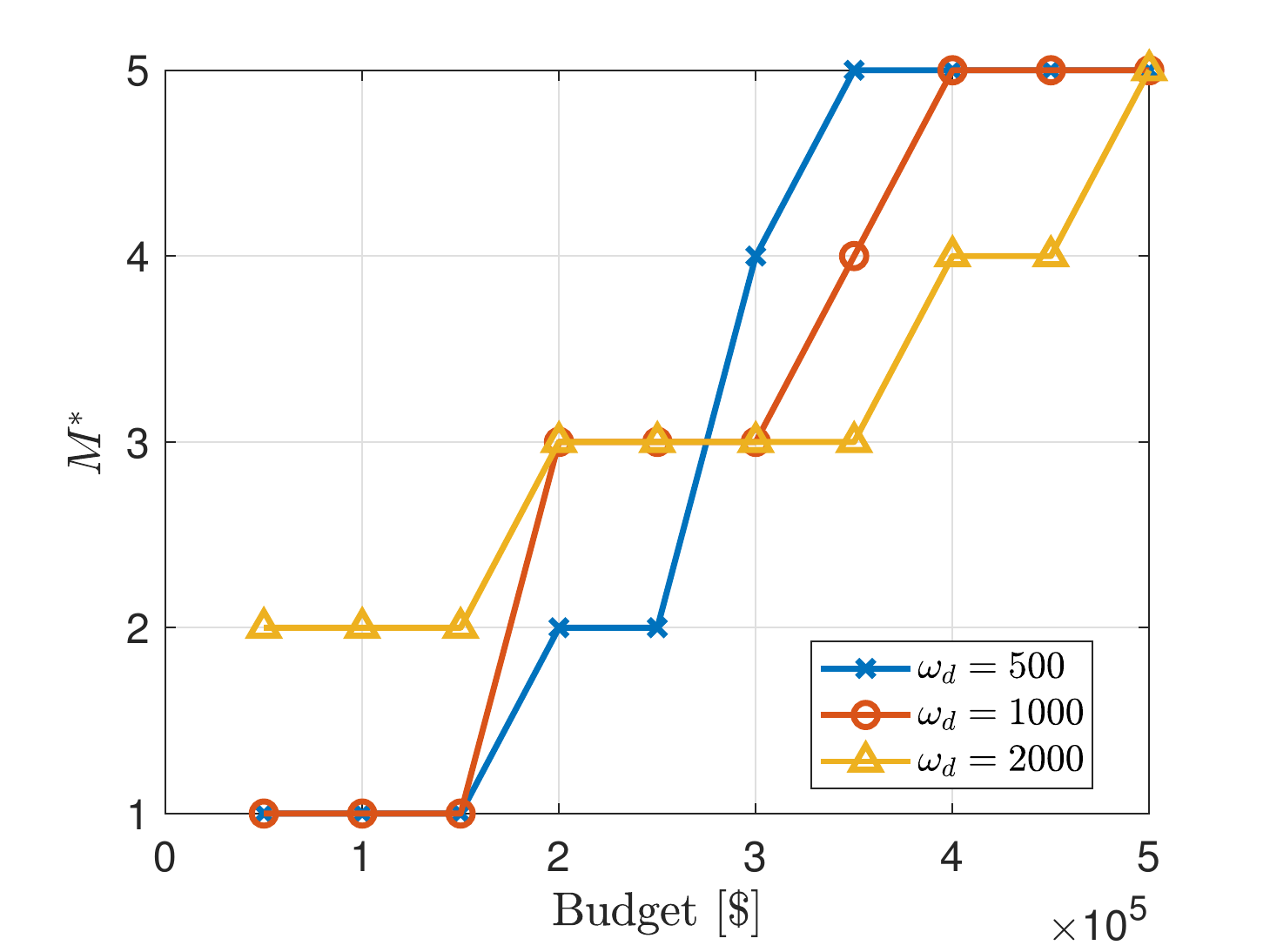}};	
		\end{tikzpicture}
	\end{subfigure}
	\caption{Optimal $\lambda_s$, $N_u$ and $M$ values for minimum wildfire losses against system budget.}  \label{fig:op2B}
\end{figure}

In Fig.~\ref{fig:op1A} and \ref{fig:op1B}, the solution of the wildfire detection maximization problem is obtained for given system budgets. As Fig.~\ref{fig:op1A} shows, the UAV-IoT system with budget $\geq 4\times 10^5  $ detects the fire with probability $>99\%$. Fig.~\ref{fig:op1B} shows the optimal solution variables. Similarly, Fig.~\ref{fig:op2A} and \ref{fig:op2B} show the solutions for the wildfire losses minimization problem. In Fig.~\ref{fig:op2A}, the minimum wildfire losses with the system budget included is shown against the system budget. The wildfire losses minimization problem is solved for three cases where the fire-related losses are modeled as $\omega_D = \omega_d \times t^2$ where $\omega_d \in \{500, 1000, 2000\} $. As the fire losses costs increase the optimal UAV-IoT system budget increases. As a result, the wildfire critical time is reduced. The stars in Fig.~\ref{fig:op2A} demonstrate the system budget at which the total wildfire losses are minimum. For instance, the minimum wildfire losses obtained by investing in the UAV-IoT system are $3.6\times 10^5, 5\times 10^5$ and $7\times 10^5$ for the cases where $\omega_d \in \{500, 1000, 2000\} $, respectively. Increasing the system budget beyond the values indicated by the stars in Fig.~\ref{fig:op2A} is not cost effective as the fire detection probability starts to saturate. Note that the cost of fire related losses if not detected by the UAV-IoT system, which means it will be detected by the satellite system, at $T_D =30$ mins, are $4.5\times 10^5, 9\times 10^5$ and $18 \times 10^5$ for the cases where $\omega_d \in \{500, 1000, 2000\} $, respectively. Thus, the UAV-IoT system can be cost effective especially for relatively high wildfire related losses costs. Finally, Fig.~\ref{fig:op2B} shows the optimal solution variables for the minimum wildfire losses problem. From this figure, we notice how it is more cost-effective to buy a UAV than to increase the IoT devices density, until some point where it becomes important to increase IoT devices density.

\section{Conclusion}
\label{sec:conclusion}

In this paper, we thoroughly proposed and analyzed the UAV-IoT system design specifically for wildfire detection purposes. We discussed the main elements of the UAV-IoT system design issues which should be optimized to achieve a desired performance. Then, we presented detailed analysis for the wildfire detection probability based on DTMC, geometry and probability theory. The analysis was verified against independent Monte Carlo simulations. Numerical results show that increasing the number of UAVs strictly improves the fire detection performance while increasing the IoT devices' density does not necessarily improve the detection probability. We also show that the UAV-IoT systems can be a cost efficient alternative to satellite imaging for wildfire detection especially when the cost of fire relevant losses is high.

\bibliographystyle{IEEEbib}
\bibliography{ref}

\end{document}